\documentclass[a4paper,11pt]{article}
\pdfoutput=1 
\usepackage{pos}

\def\eq#1{(\ref{#1})}
\def\beqs{\begin{eqnarray}}
\def\eeqs{\end{eqnarray}}
\newcommand{\mathd}{\mathrm{d}}
\newcommand{\mathe}{\mathrm{e}}

\title{Emergence of classical spacetimes in canonical tensor model}

\author*[a]{Naoki Sasakura}

\affiliation[a]{Yukawa Institute for Theoretical Physics, Kyoto University,\\
  Oiwake, Kitashirakawa, Sakyo-ku, Kyoto, 606-8502, Japan}

\emailAdd{sasakura@yukawa.kyoto-u.ac.jp}

\abstract{
We study the wave function of a tensor model in the canonical formalism by Hamiltonian Monte Carlo method
for Lie group symmetric or nearby values for the argument of the wave function,
and show that there emerge Lie-group symmetric semi-classical spacetimes. 
More precisely, we consider some $SO(n+1)\ (n=1,2,3)$ symmetric values for the tensor argument of the wave function, and 
show that there emerge discrete $n$-dimensional spheres.   
A key fact is that there exist two phases, the classical phase and the quantum phase, depending
on the values of the argument of the wave function,
and emergence of classical spaces above occurs in the former phase, 
while fluctuations of configurations are too large for such emergence in the latter phase. 
The transition between the two phases has similarity with the Gross-Witten-Wadia transition, or that between the one-cut
and the two-cut solutions in the matrix model. 
Based on the results, we give some speculations on how spacetimes evolve in the tensor model.
}

\FullConference{%
  Corfu Summer Institute 2021 "School and Workshops on Elementary Particle Physics and Gravity"\\
  29 August - 9 October 2021\\
  Corfu, Greece
}

\notes{\note{Preprint:  {\sf YITP-22-31}}}

\begin{document}

\maketitle

\section{Introduction}
\label{sec:introduction}
Planck length is considered to be the length scale where quantum gravitational aspects become evident \cite{Garay:1994en}.
While it is not certain whether this length scale should be associated to a discrete nature of spacetime,
discrete approaches to quantum gravity have some advantages, such as enabling us to leave from 
continuum spacetime picture at the fundamental level, 
which usually requires some predetermined assumptions such as smoothness and dimensions.
On the other hand, from the perspective of discrete spacetime, continuum picture of spacetime is an effective description 
which emerges in the infrared, resulting from dynamics.

One of the discrete approaches to quantum gravity is given by the tensor model \cite{Ambjorn:1990ge,Sasakura:1990fs,Godfrey:1990dt,Gurau:2009tw}. 
The tensor model was introduced to generalize the matrix model, which is successful in describing
two-dimensional quantum gravity, to dimensions more than two. 
Unfortunately the tensor model suffers from the dominance of singular spaces (like branched polymers \cite{Bonzom:2011zz,Gurau:2011xp})
and has not been successful in generating macroscopic spaces.
An interesting possibility to improve the situation would be to introduce a temporal direction to the tensor model.
This is suggested by the following fact which implies the importance of a temporal direction in quantum gravity:
It has  been shown that the causal dynamical triangulation (CDT),
which is a discrete model of quantum gravity with a temporal direction, is successful in generating macroscopic
spacetimes \cite{Ambjorn:2004qm}, while the dynamical triangulation (DT), which is Euclidean, is not.

Generally, delicate treatment is necessary in introducing a temporal direction to a discrete model of quantum gravity. 
The reason is that the expected infrared effective theory, namely general relativity, is invariant under temporal 
diffeomorphism (more thoroughly, spacetime diffeomorphism), 
and therefore a fundamental discrete theory has to contain a mechanism which assures the invariance in the infrared. 
To incorporate this requirement, the present author formulated a tensor model, which
we call the canonical tensor model (CTM) \cite{Sasakura:2011sq,Sasakura:2012fb}, as a first-class 
constraints system in the Hamiltonian formalism, mimicking the Hamiltonian
formalism of general relativity (more precisely, the ADM formalism \cite{Arnowitt:1962hi}). 

Though the dynamical variables are largely different between CTM and general relativity (GR),
the similarity between their formalisms leads to some explicit connections between them. 
The $N=1$ case\footnote{The dynamical variables of CTM
are a canonically conjugate pair of real symmetric order-three tensors, $Q_{abc},\ P_{abc}\ (a,\cdots=1,2,\cdots,N)$.} 
of CTM has been shown to be equivalent to a mini-superspace treatment of GR \cite{Sasakura:2014gia}.
In a formal continuum limit\footnote{This is analogous to a vanishing limit of a lattice interval in a lattice theory without
taking into account dynamics. Dimension of spacetime is assumed as an input.}, 
the constraint algebra of CTM has been shown to agree with that of ADM \cite{Sasakura:2015pxa}.
Also in the formal continuum limit, it has been shown that
the equation of motion of CTM agrees with a Hamilton-Jacobi equation of GR with a Hamilton's principal function
of a local form containing also a scalar and higher spin fields \cite{Chen:2016ate}. 

The quantization of CTM can be carried out by simply applying the canonical quantization \cite{Sasakura:2013wza}. A convenient
fact is that the closure of the constraint algebra holds even after quantization, namely, the 
first class nature is kept after quantization. Then the physical state condition
can be consistently imposed by requiring the quantized constraints to vanish on the physical states. 
The condition is represented by a system of non-linear partial differential equations for the wave function.
A surprising fact is that, though the system of partial differential equations is complicated, there exists an exact solution for general $N$,
which has the form of a multi-variable generalization of the Airy function \cite{Narain:2014cya}.

The study of the wave function will reveal the quantum nature of CTM, possibly leading to the emergence of 
spacetime in CTM. An important previous result is that the wave function has peaks 
at Lie-group invariant values of its tensor argument \cite{Obster:2017dhx}, the mechanism of which can be readily understood
as a quantum coherence phenomenon \cite{Obster:2017pdq}. This fact physically means that Lie group symmetries emerge in CTM.
Considering that spacetime structure can be determined by some Lie group symmetries, such as Poincare or de Sitter
symmetries, it would be possible that emergence of spacetime could be realized by the same mechanism as the 
emergence of Lie group symmetries.

This question was pursued in the last two works \cite{Sasakura:2021lub,Kawano:2021vqc}. 
It has been found that there exist two phases, which we respectively call
the quantum phase and the classical phase \cite{Kawano:2021vqc}, 
depending on the values of the argument of the wave function.
While the configurations fluctuate largely in the quantum phase, 
the fluctuation is suppressed in the classical phase and classical spacetimes emerge.
More precisely, we took $SO(n+1)\ (n=1,2,3)$ symmetric values for the tensor argument of the wave function and have found that
there emerge discrete $n$-dimensional spheres. 
The main purpose of this paper is to review this development in the previous paper \cite{Kawano:2021vqc}.

In Section~\ref{sec:CTM}, we review the formulation of CTM. 
In Section~\ref{sec:HMC}, we explain the setup of our Hamiltonian Monte Carlo
simulation with the reweighting method. In Section~\ref{sec:qc}, we show the presence of two phases, the quantum and the classical phases.
In Section~\ref{sec:emergence}, we show the emergence of classical spacetimes in the classical phase. More precisely, 
we observe emergence of discrete $n$-dimensional spheres for $SO(n+1)\ (n=1,2,3)$ invariant values of the argument. 
In Section~\ref{sec:different}, we add perturbations to the symmetric values of the argument or change the representations, 
and show that the classical phase becomes less likely under the perturbations or the changes.
In Section~\ref{sec:speculations}, we speculate the spacetime evolution in CTM. 
The last section is devoted to summary and future prospects.
 
\section{Canonical tensor model (CTM)}
\label{sec:CTM}
The canonical tensor model is a tensor model in the Hamiltonian formalism. This is formulated as a first-class constraints system
\cite{Sasakura:2011sq} mimicking the Hamiltonian formalism of general relativity, more precisely, the ADM formalism \cite{Arnowitt:1962hi}.
In the classical case, the dynamical variables are a canonically conjugate pair of real symmetric order-three tensors, $Q_{abc}$ and $P_{abc}\ (a,b,c=1,2,\ldots,N)$,
which satisfy the fundamental Poisson brackets,
\begin{eqnarray}
&&\{ Q_{abc}, P_{def} \}=\sum_\sigma \delta_{a\sigma_d} \delta_{b\sigma_e} \delta_{a\sigma_f}, \cr
&&\hbox{Others}=0,
\end{eqnarray}
where the sum is over all the permutations of $d,e,f$.
The Hamiltonian is given by a linear combination,
\beqs
H=N_a {\cal H}_a +N_{[ab]} {\cal J}_{[ab]},
\eeqs
where $N_a$ and $N_{[ab]}$ are respectively the analogues of the lapse and the shift in ADM, and ${\cal H}_a $ and ${\cal J}_{[ab]}$
are respectively the analogues of the Hamiltonian and the momentum constraints in ADM. Here the bracket $[\cdot]$ 
in the indices represents the anti-symmetry, $N_{[ab]}=-N_{[ba]},\ {\cal J}_{[ab]}=-{\cal J}_{[ba]}$.

As in ADM, the ${\cal H}_a$ and ${\cal J}_{[ab]}$ are required to form a closed Poisson algebra, namely, they are required to be
first-class constraints.
This requirement of the Poisson algebraic closure is so strong that, with some additional physically reasonable conditions, 
the constraints are uniquely determined \cite{Sasakura:2012fb} to be
\begin{eqnarray}
{\cal H}_a&=& \frac{1}{2}\left( P_{abc}P_{bde} Q_{cde} -\Lambda \, Q_{abb} \right), \cr
{\cal J}_{[ab]}&=& \frac{1}{4} \left(P_{acd}Q_{bcd} -P_{bcd}Q_{acd} \right),
\label{eq:handl}
\end{eqnarray}
where $\Lambda$ is a real constant. 

In the $N=1$ case, CTM agrees with a mini-superspace treatment of GR \cite{Sasakura:2014gia}, in which
$\Lambda$ corresponds to the cosmological constant. 
Therefore we call $\Lambda$ the cosmological
constant of CTM. Since the Poisson algebraic structure does not change under the rescaling $Q\rightarrow s Q,\ P\rightarrow P/s$
with an arbitrary real $s$ and $\Lambda$ is effectively changed by $s^2 \Lambda$ under the rescaling, $\Lambda$ can 
be normalized to be, for instance, $\Lambda=0,\pm 1$. In this paper, we consider only the case of a positive cosmological
constant, and normalize it as $\Lambda=4/9$ for later convenience of our Monte Carlo simulation.

Let us explain the reason why we consider only the case of a positive cosmological constant. In Section~\ref{sec:introduction}, 
we explained that the wave function (in \eq{eq:varphip} below) of CTM has the property that peaks appear at 
Lie-group symmetric values of the tensor argument \cite{Obster:2017dhx}. 
This phenomenon is actually evident only for the case of a positive cosmological constant.   
This means that the emergence of Lie group symmetries in CTM appears only for a positive cosmological constant. 
Therefore, considering the expected link between the emergence of Lie group symmetries and that of spacetimes, it would 
be reasonable to restrict our interest to the case of a positive cosmological constant. 
In fact, in \cite{Sasakura:2019hql,Obster:2020vfo}, 
the negative case was studied by approximating the Airy function part (discussed later) of the wave function by
a Gaussian function, and emergence of spacetimes was not found.

From \eq{eq:handl} the Poisson algebra of ${\cal H}_a$ and ${\cal J}_{[ab]}$ is given by
\begin{eqnarray}
&&\{ {\cal H}\xi^1,{\cal H}\xi^2\}={\cal J}([P\xi^1,P\xi^2]+2 \Lambda [\xi^1 ,\xi^2]),\cr
&&\{ {\cal J}\eta,\, {\cal H}\xi\}={\cal H}(\eta \xi),\cr
&&\{ {\cal J}\eta^1,\,{\cal  J}\eta^2\}={\cal J}([\eta^1,\eta^2]), 
\label{eq:algctm}
\end{eqnarray}
where we have introduced auxiliary c-number variables $\xi_a,\eta_{[ab]}$, 
\begin{eqnarray}
&&{\cal H}\xi:={\cal H}_a\xi_a,\ {\cal J}\eta:= {\cal J}_{[ab]}\eta_{[ab]},\ (P\xi)_{ab}:=P_{abc}\xi_c,\cr
&&[\xi^1,\xi^2]_{ab}:= \xi^1_a \xi^2_b-\xi^1_b \xi^2_a,\ (\eta\xi)_a:= \eta_{[ab]}\xi_b,
\end{eqnarray}
for simplicity of expressions, and $[P\xi^1,P\xi^2]$ in the first line represents the matrix commutator.
 
The quantization of CTM can be carried out by the canonical quantization. The dynamical variables are now promoted 
to the quantum ones with
\begin{eqnarray}
&&[\hat Q_{abc},\hat P_{def} ]=i \sum_\sigma \delta_{a\sigma_d} \delta_{b\sigma_e} \delta_{a\sigma_f}, \cr
&&\hbox{Others}=0,
\label{eq:funpoi}
\end{eqnarray}
where $\hat Q_{abc},\hat P_{def}$ are again symmetric tensors and assumed to be hermite. 
The quantized constraints are given by
\begin{eqnarray}
&&\hat {\cal H}_a= \frac{1}{2}\left( \hat P_{abc} \hat P_{bde} \hat Q_{cde} + 2 i R \hat P_{abb} -\Lambda \hat Q_{abb} \right), \cr
&&\hat {\cal J}_{[ab]}= \frac{1}{4} \left(\hat P_{acd} \hat Q_{bcd} -\hat P_{bcd} \hat Q_{acd} \right).
\label{eq:qhandl}
\end{eqnarray}
Here the only difference from the classical ones \eq{eq:handl} is the presence of the middle term in $\hat{\cal H}$, which
realizes the hermiticity of  $\hat{\cal H}$ by taking
\beqs
R=\frac{(N+2)(N+3)}{4}.
\label{eq:valueR}
\eeqs
A convenient fact is that, by explicit computation, the commutation algebra of the quantized constraints closes as 
\begin{eqnarray}
&&[ \hat {\cal H}\xi^1,\hat{\cal H}\xi^2]=i \hat {\cal J}([\hat P\xi^1,\hat P\xi^2]+2 \Lambda [\xi^1 ,\xi^2]),\cr
&&[ \hat {\cal J}\eta,\, \hat {\cal H}\xi]=i \hat {\cal H}(\eta \xi),\cr
&&[ \hat  {\cal J}\eta^1,\hat {\cal  J}\eta^2 ]=i \hat {\cal J}([\eta^1,\eta^2]), 
\label{eq:qalgctm}
\end{eqnarray}
where\footnote{The algebra \eq{eq:qalgctm} consistently holds also 
for $\hat {\cal H}\xi:=\xi_a \hat {\cal H}_a$, etc., in which the auxiliary parameters
appear in the opposite side.
This can be shown by taking the hermitian conjugate of the algebra. 
This reordering is relevant on the righthand side of the first equation, in which the argument contains the operator $\hat P$.
With this reordered expression, the physical state condition \eq{eq:psc} can be more directly derived. }
\begin{eqnarray}
&&\hat {\cal H}\xi:=\hat {\cal H}_a\xi_a,\ \hat {\cal J}\eta:= \hat {\cal J}_{[ab]}\eta_{[ab]},\ (\hat P\xi)_{ab}:=\hat P_{abc}\xi_c,\cr
&&[\xi^1,\xi^2]_{ab}:= \xi^1_a \xi^2_b-\xi^1_b \xi^2_a,\ (\eta\xi)_a:= \eta_{[ab]}\xi_b.
\end{eqnarray}
Namely, the first-class nature of CTM does not change after quantization. Therefore, one can consistently impose 
\beqs
\hat{\cal H}_a | \Psi\rangle=\hat {\cal J}_{[ab]} |\Psi \rangle=0,
\label{eq:psc}
\eeqs
as the physical state condition.

From the form \eq{eq:qhandl}, the physical state condition \eq{eq:psc} for the wave function corresponding to $|\Psi\rangle$ 
becomes a system of partial differential equations non-linear in $Q,P$. A surprising fact is that, in spite of the non-linear
structure, there exits an exact solution valid for general $N$. This is given in $P$-representation by
\beqs
&&\Psi(P):=\langle P | \Psi \rangle=\varphi(P)^R,
\label{eq:power}
\eeqs
where 
\beqs
\varphi(P):=\int \prod_{a=1}^N \mathd\phi_a \, \mathd\tilde \phi\, \mathe^{i \left(P\phi^3-\phi^2 \tilde \phi 
+\frac{4}{27 \Lambda} \tilde \phi^3 \right)}
\label{eq:varphip}
\eeqs
with
\beqs
&&P\phi^3:=P_{abc}\phi_a\phi_b \phi_c,   \cr
&&\phi^2:=\phi_a \phi_a.
\eeqs
If $R$ is an integer, the expression \eq{eq:power} with the power $R$ can be equivalently rewritten by introducing $R$ replicas of variables 
as  
\beqs
\Psi(P)= \int_{\cal C} \mathd\phi\, \mathd\tilde \phi\, \mathe^{i\sum_{j=1}^R \left(P\phi^j{}^3-\phi^j{}^2 \tilde \phi^j 
+\frac{4}{27 \Lambda} \tilde \phi^j{}^3 \right)},
\label{eq:psip}
\eeqs
where the integration variables are $\phi_a^j,\tilde \phi^j\ (a=1,2,\ldots,N,\ j=1,2,\ldots,R)$, and 
\beqs
&&\mathd\phi\, \mathd\tilde \phi:=\prod_{j=1}^R  \mathd\tilde \phi^j  \prod_{a=1}^N  \mathd\phi_a^j.
\eeqs
Here $R$ plays the role of a replica number.
The integration contour ${\cal C}$ in \eq{eq:psip} should be chosen so that the integration converge. The most rigorous way
to define such a contour is to consider a Lefschetz thimble \cite{Witten:2010cx}. It is, however, also possible to take it as  
${\cal C}=(\mathbb{R}^N\times \tilde {\cal C})^R$, 
where $\mathbb{R}^N$ is for each $\phi^j$ and $\tilde {\cal C}$ for each $\tilde \phi^j$, by additionally introducing an infinitesimal 
regularization term for the convergence \cite{Obster:2017dhx}.
$\tilde {\cal C}$ will be specified in Section~\ref{sec:HMC}.  

\section{Hamiltonian Monte Carlo setup}
\label{sec:HMC}
The wave function $\Psi(P)$ in \eq{eq:psip} shows the phenomenon that it has peaks at Lie group symmetric values of $P$. 
It is plausible that this symmetry emergence phenomenon also triggers the emergence of spacetimes, since they are 
characterized by de Sitter or Poincare symmetries. What has been found in the previous work \cite{Kawano:2021vqc}
is that emergence of spacetimes can indeed be found for the wave function $\Psi(Q)$  
in the $Q$-representation of the state $|\Psi\rangle$ rather than in $\Psi(P)$ above.

Let us define $\Psi(Q)$ by the Fourier transform of $\Psi(P)$ as\footnote{
To obtain the $Q$-representation exactly consistent with the normalization employed in \eq{eq:funpoi}, 
the Fourier transform should be taken with a kernel $\mathe^{i Q_{abc} P_{abc}/6}$ rather than \eq{eq:foupsi}.
However, this numerical difference does not change the essentials, and we take \eq{eq:foupsi} for the simplicity of the expression. 
The actual $Q$-representation can be recovered by $Q\rightarrow Q/6$.}
\beqs
\Psi(Q):=\int_{\mathbb{R}^{\# P}} \prod_{d\leq e \leq f =1}^N \mathd P_{def}\, \mathe^{i Q_{abc}P_{abc}}\, \Psi(P),
\label{eq:foupsi}
\eeqs
where $\# P= N(N+1)(N+2)/6$, the number of independent components of $P$.
Since the exponent in \eq{eq:psip} is linear in $P$, the Fourier transform \eq{eq:foupsi} results in a product of a number of 
Dirac $\delta$-functions, which is difficult to be treated by Monte Carlo simulations.  Therefore we consider a smeared wave function,
\beqs
\Psi(Q,\lambda)&:=&{\rm const.} \int_{\mathbb{R}^{\# Q} }\prod_{a\leq b \leq c =1}^N \mathd \tilde Q_{abc} \, 
\mathe^{-\lambda (Q-\tilde Q)^2} \Psi(\tilde Q) \cr
&=& {\rm const.} 
 \int_{\cal C} \mathd\phi\, \mathd\tilde \phi\, \mathe^{
 -\lambda (Q-\phi \phi \phi)^2+
 i\sum_{j=1}^R \left(-\phi^j{}^2 \tilde \phi^j 
+\frac{4}{27 \Lambda} \tilde \phi^j{}^3 \right)},
\label{eq:psiqlam}
\eeqs
where $\lambda$ is positive real, $Q^2=Q_{abc}Q_{abc}$, $\# Q=\# P$, and $(\phi \phi \phi)_{abc}:=\sum_{j=1}^R \phi_a^j
 \phi_b^j \phi_c^j$. Here we have performed $\phi\rightarrow -\phi$ for convenience.
 The original wave function can be recovered in the zero smearing limit, $\lambda \rightarrow \infty$.
 In our Monte Carlo simulation, we take it as large as $\lambda\leq 10^7$.

The integrand in \eq{eq:psiqlam} is complex and therefore suffers from the notorious sign problem \cite{Berger:2019odf}. 
To deal with this issue, we apply the so-called 
reweighting method. With this method, the real part defines the system for the Monte Carlo simulation, while 
the complex part is computed as an expectation value of an observable in the system. 
In our case, the wave function is expressed as
\beqs
\Psi(Q,\lambda)= {\rm const.}\, Z_{Q,\lambda}\, \left \langle \prod_{j=1}^R {\rm Airy}(-\phi^j{}^2) \right \rangle_{Q,\lambda},
\label{eq:psiqmonte}
\eeqs
where we have taken $\Lambda=4/9$ for convenience 
(without losing generality for the positive case, as explained in Section~\ref{sec:CTM}), have used the integral expression 
of the Airy function,
\beqs
{\rm Airy}(-z)={\rm const.} \int_{\tilde C} \mathd \tilde \phi \, \mathe^{i \left(-z \tilde \phi+\frac{1}{3} \tilde \phi^3\right)},
\eeqs
and  
\beqs
Z_{Q,\lambda}:=\int_{\mathbb{R}^{NR}}\mathd\phi  
\, \mathe^{ -\lambda (Q-\phi \phi \phi)^2}.
\label{eq:part}
\eeqs
In \eq{eq:psiqmonte}, $\left \langle \cdot \right \rangle_{Q,\lambda}$ represents an expectation value
in the system defined by the partition function \eq{eq:part}.

\begin{figure}
\begin{center}
\includegraphics[width=5cm]{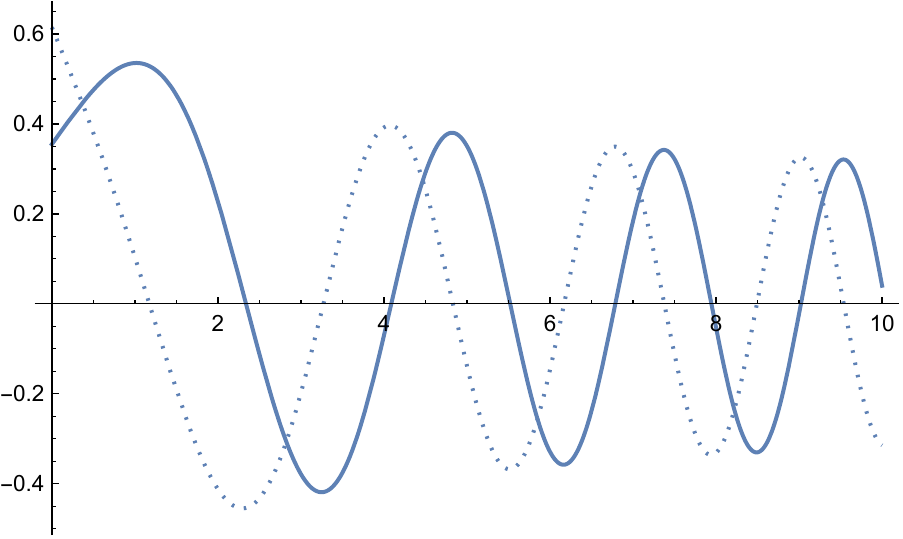}
\hfil
\includegraphics[width=3.5cm]{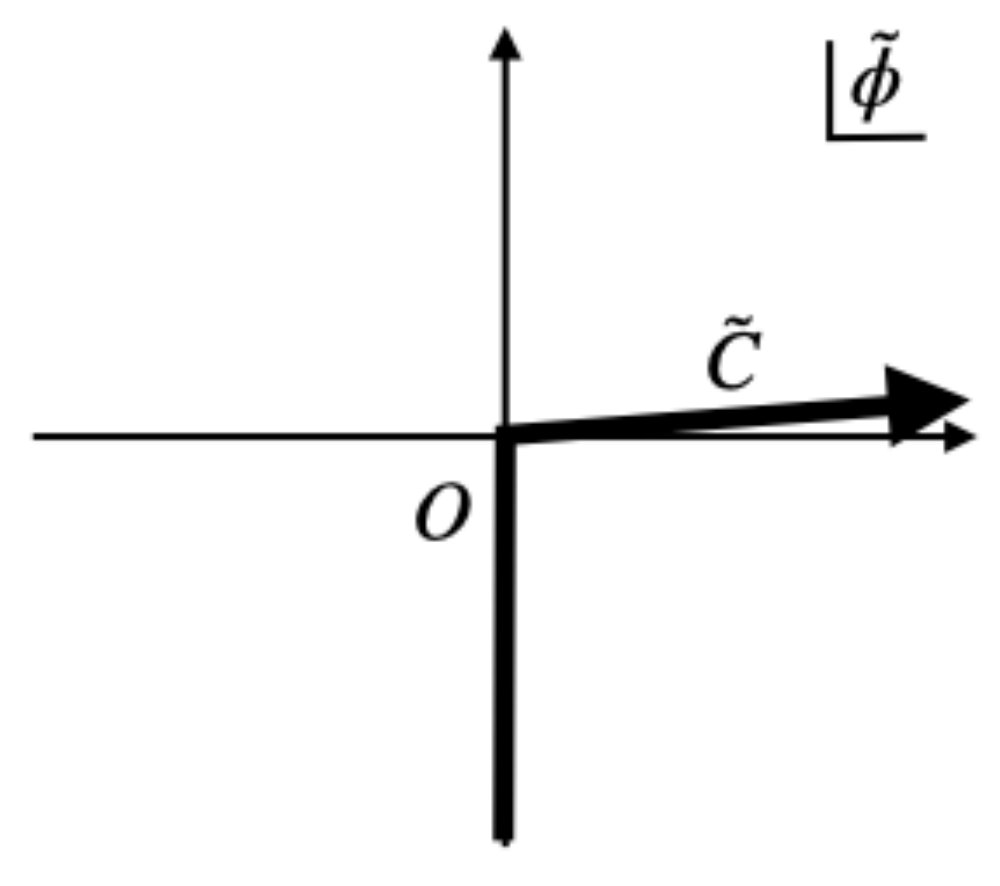}
\caption{Left: Real and imaginary parts of \eq{eq:airy} plotted against $z>0$. It is oscillatory.
Right: The contour $\tilde C$ for \eq{eq:airy}.}
\label{fig:airy}
\end{center}
\end{figure}

As for the integration contour $\tilde C$, which stretches to infinity, there are two independent choices of $\tilde C$, which generate 
the two independent Airy functions, ${\rm Ai}(-z)$ and ${\rm Bi}(-z)$, respectively. We take $\tilde C$ so that 
the Airy function in \eq{eq:psiqmonte} be
\beqs
{\rm Airy}(-z)={\rm Ai}(-z)+ i\, {\rm Bi} (-z),
\label{eq:airy}
\eeqs
up to an overall factor.
This choice of Airy function has the form of a propagating wave $\sim \mathe^{-2 i  z^{3/2}/3}$ in the asymptotic
limit $z\rightarrow \infty$ (See Figure~\ref{fig:airy}). 
This choice would be a natural one in the sense that it would represent a spacetime evolving in one direction; 
an expanding (or contracting) spacetime\footnote{ 
It is also possible to chose for instance Ai or Bi instead, which are real. In this case, the wave function will represent
a standing wave, which can be thought to be a superposition of an expanding and a contracting spacetime.}.  

In fact, whether the partition function \eq{eq:part} converges or not is a non-trivial question, because there exist flat directions 
extending to infinity of $\phi$. A simple example of a flat direction is the one which has a pair of $\phi^i=-\phi^j$. 
In this case, the summation $\phi \phi \phi$ in \eq{eq:part} contains a part, $\phi_a^i \phi^i_b \phi^i_c+ \phi_a^j \phi^j_b \phi^j_c=0$, 
whatever the size of $\phi^i$ and $\phi^j$ is,  and therefore the flat direction extends to infinity. 
This means that the integrand does not damp exponentially toward infinity in such directions, and 
the convergence can be guaranteed only by the property of the integration volume.
The question of the convergence of \eq{eq:part} was studied systematically in \cite{Obster:2021xtb} with assistance of numerics. 
The conclusion/conjecture is that the integral converges for $R<(N+1)(N+2)/2$.  For $N>1$, this condition 
is satisfied for \eq{eq:valueR}, and the MC simulation should have no problems. Indeed we did not encounter any divergent behavior
in our simulation.

The expression \eq{eq:psip} is valid only for integer $R$, meaning that $N$ is restricted due to \eq{eq:valueR}.
This restriction of $N$ is inconvenient, so we take
\beqs
R=\left \lfloor \frac{(N+2)(N+3)}{4} \right \rfloor
\label{eq:floorR}
\eeqs
instead, where $\lfloor \cdot \rfloor$ is the floor function, to allow us to take $N$ freely in our Monte Carlo simulation.
This is based on the assumption that the small change of the value of $R$ does not 
affect the dynamics of the system \eq{eq:part}  in an essential manner.

We performed Hamiltonian Monte Carlo simulation \cite{Neal(2011)} to study the system \eq{eq:part}. 
The leapfrog numbers were typically taken with a few hundreds
depending on the sizes of $N,R$. 
The total numbers of the samples in each sequence were typically around $10^4\sim 10^6$, 
and one data per $\sim 10^2$ raw data was used for computations of observables to remove correlations.
The machine had Xeon W2295 (3.0GHz, 18 cores), 128GB DDR4 memory, and Ubuntu 20 as OS.  
The program was written in C++ with pthread for parallelization. As for the Airy function, the boost 
library \cite{boost} was used. Every run typically took several hours with active use of parallelization.

\section{Quantum and classical phases}
\label{sec:qc}
The properties of the wave function in \eq{eq:psiqmonte} will depend on the dynamics of the system defined by \eq{eq:part}, 
which will non-trivially depend on the values of $Q$.
Since $Q$ has so many components of order $\sim N^3/6$, it is currently not possible to state the dynamics
for general values of $Q$. Rather, to be concrete in this paper, 
we will restrict ourselves to considering only specific values of $Q$ invariant under $SO(n+1)\ (n=1,2,3)$, 
and some perturbations from these values, with specific physically natural choices of representations. 
As we will see in due course, these limited cases are still very interesting: There exist two phases, the quantum and the classical, and 
there emerge discrete $n$-dimensional spheres in the classical phase.

Let us show the construction of $Q$. Consider an $n$-dimensional sphere $S^n$, and the harmonic
functions on it. For $n=1$ case, a set of normalized harmonic functions $\{f_a\, |\ a=1,2,\ldots,N\, (=2M+1)\}$ are given by
\beqs
\{f_a\}=\left \{ \frac{1}{\sqrt{2}}  \right\}  \cup  \left\{ \cos(m \theta),\ \sin(m \theta)\  | \ m=1,2,\ldots, M \right\},
\eeqs
where $\theta \in [0,2 \pi )$ is the coordinate on $S^1$. $M$ is physically a momentum cutoff, and 
the total number of the harmonic functions is $N=2M+1$. Then an $SO(2)$-invariant tensor can be obtained by 
\beqs
Q^{SO(2)}_{abc}=\hbox{const.}\, e^{-\alpha (m_a^2+m_b^2+m_c^2)/M^2 }\int_0^{2 \pi} \mathd \theta\, f_a f_b f_c,
\label{eq:qso2}
\eeqs
where $\alpha$ is a positive number 
and $m_a$ are the momentums associated to $f_a$ (for instance, $m_a=m$ for $f_a=\cos(m \theta)$).
The role of this dumping factor is to moderate the momentum cutoff $M$. If this moderation is not taken, 
$M$ becomes a sharp cutoff and will generate non-local behavior in $\theta$, 
which would be physically unwelcome\footnote{This very common aspect can be explicitly seen by comparing, for example, $\sum_{m=-M}^M \mathe^{i m \theta}$ and $\sum_{m=-M}^M 
\mathe^{-\alpha m^2+i m \theta}$. The former function is a non-local oscillating function, while the latter moderated one
has a peak concentrated around $\theta \sim 0$. }.
The overall constant in \eq{eq:qso2} is taken so that $Q$ is normalized\footnote{The norm of a tensor is defined by
$|Q|=\sqrt{Q_{abc}Q_{abc}}$.} by $|Q|=1$.
It is elementary to understand that $Q_{abc}$ is invariant under $SO(2)$ transformation, which rotates $\theta$, generating 
the charge-$m$ representation on the space spanned by $\cos(m \theta)$ and $\sin(m \theta)$.

The above procedure can be extended to any $n$. For $S^2$, we take
\beqs
\{f_a\}=\{ {\rm Re}\, Y_l^m(\Omega),\ {\rm Im}\,Y_l^m(\Omega)\ | \ m=-l,-l+1,\ldots, l,\ l=0,1,\ldots,L\},
\eeqs
where $\Omega$ is a coordinate on $S^2$, and $Y_l^m$ are the spherical harmonics. Here, it is implicitly assumed that
only independent ones are included (For instance, ${\rm Im}\,Y_L^0=0$ is discarded).
The total number of the functions is $N=(L+1)^2$. An $SO(3)$-invariant tensor is given by 
\beqs
Q^{SO(3)}_{abc}=\hbox{const.}\, e^{-\alpha (l_a^2+l_b^2+l_c^2)/L^2 }\int_{S^2}\mathd\Omega \, f_a f_b f_c,
\label{eq:qso3}
\eeqs
which is invariant under the $SO(3)$ rotation on $S^2$, generating the spin $l$ representation on $Y_l^m\ (m=-l,-l+1,\ldots,l)$.
The construction for any $S^n\ (n>2)$ is basically the same with the generalized spherical harmonics \cite{Kawano:2021vqc}.

\begin{figure}
\includegraphics[width=3.5cm]{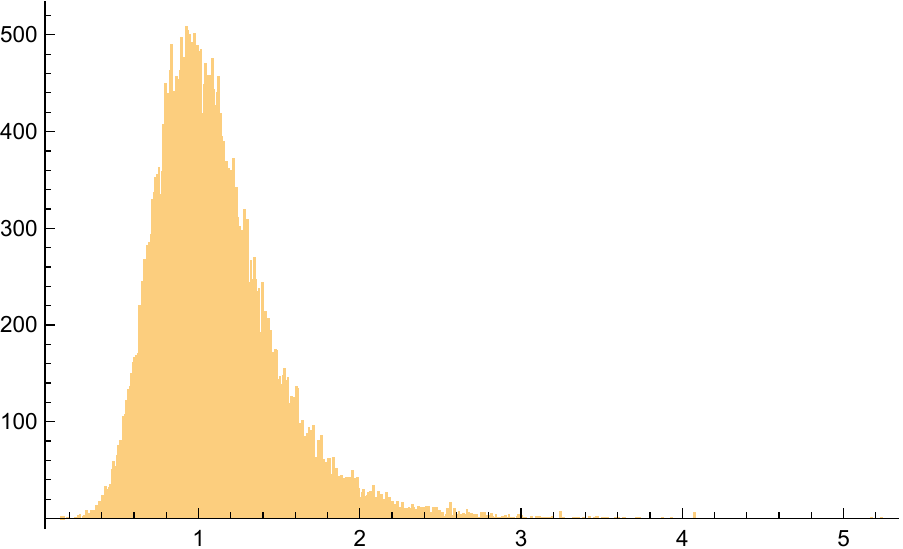}
\hfil
\includegraphics[width=3.5cm]{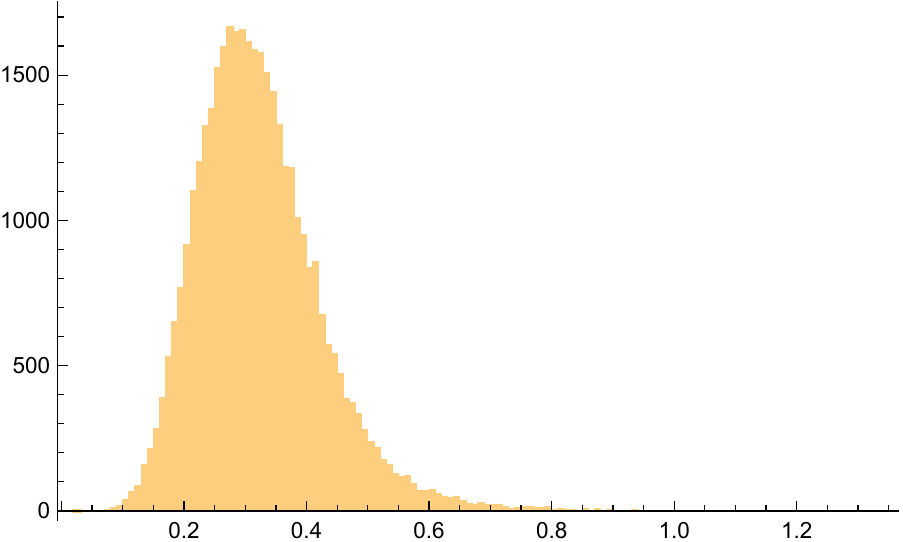}
\hfil
\includegraphics[width=3.5cm]{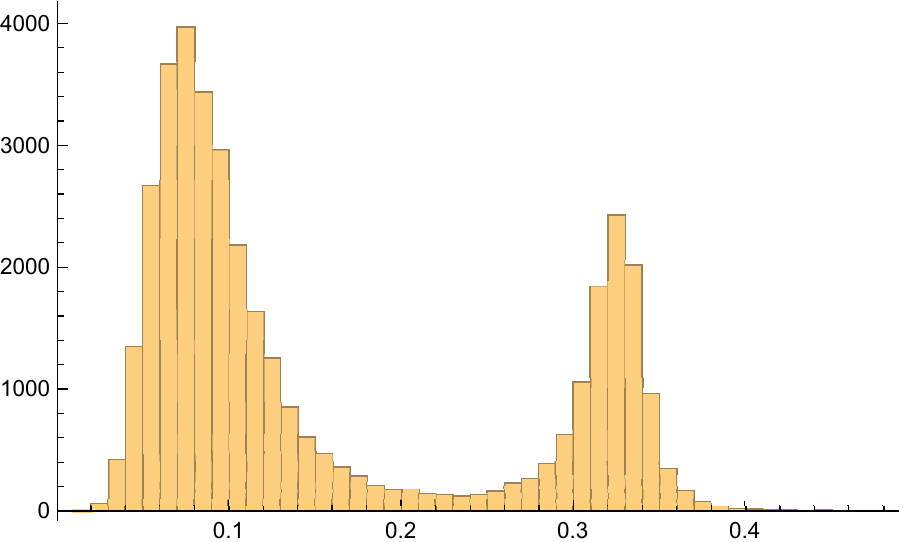}
\hfil
\includegraphics[width=3.5cm]{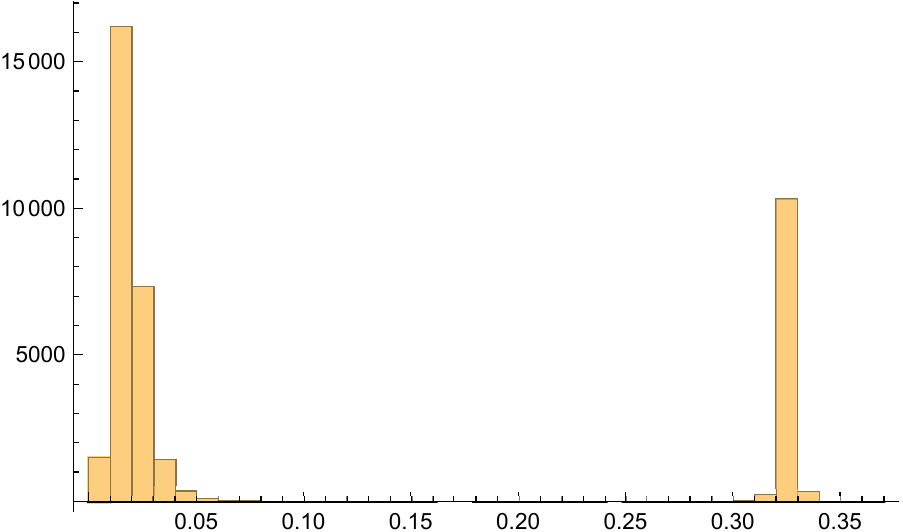}
\caption{The histograms of $\phi^j{}^2$ for $Q=Q^{SO(2)}$ with $N=15\ (M=7)$, $R=76$, 
and $\alpha=0.5$. The values 
of $\lambda$ are $10$, $10^3$, $10^5$, and $10^7$, respectively from the left to the right. There is a transition of 
the topology of the distribution.}
\label{fig:transN15}
\end{figure}

\begin{figure}
\includegraphics[width=3.5cm]{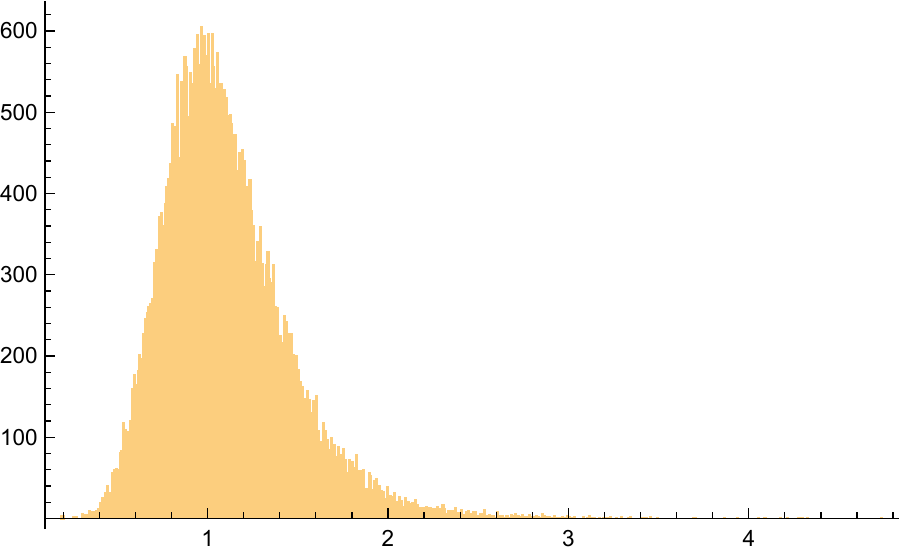}
\hfil
\includegraphics[width=3.5cm]{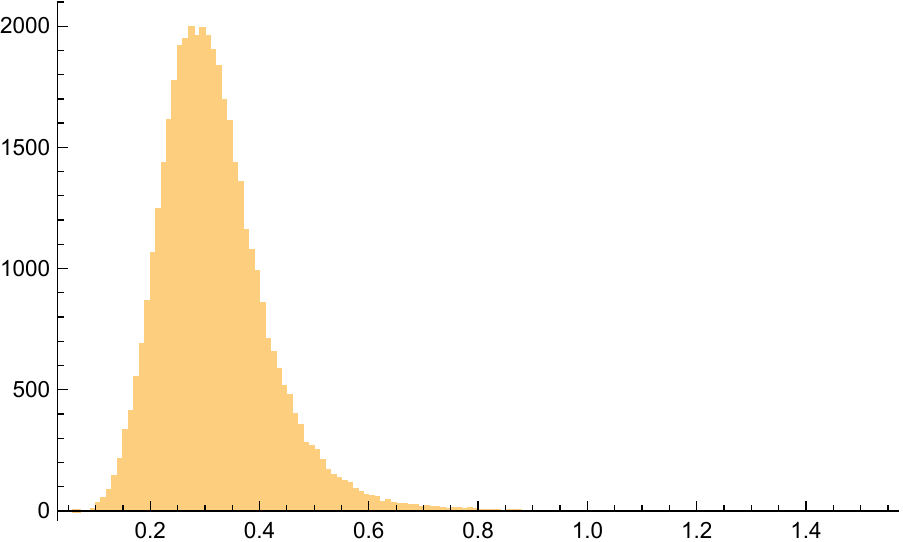}
\hfil
\includegraphics[width=3.5cm]{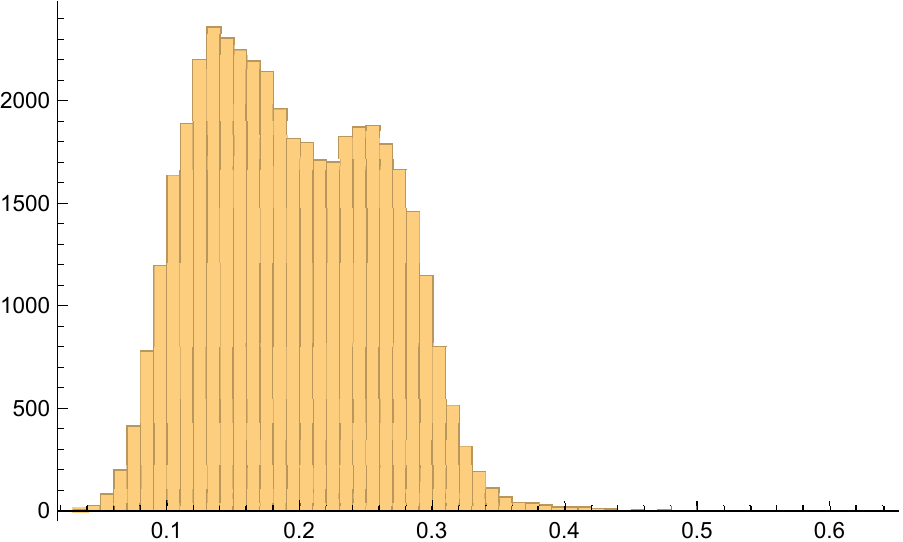}
\hfil
\includegraphics[width=3.5cm]{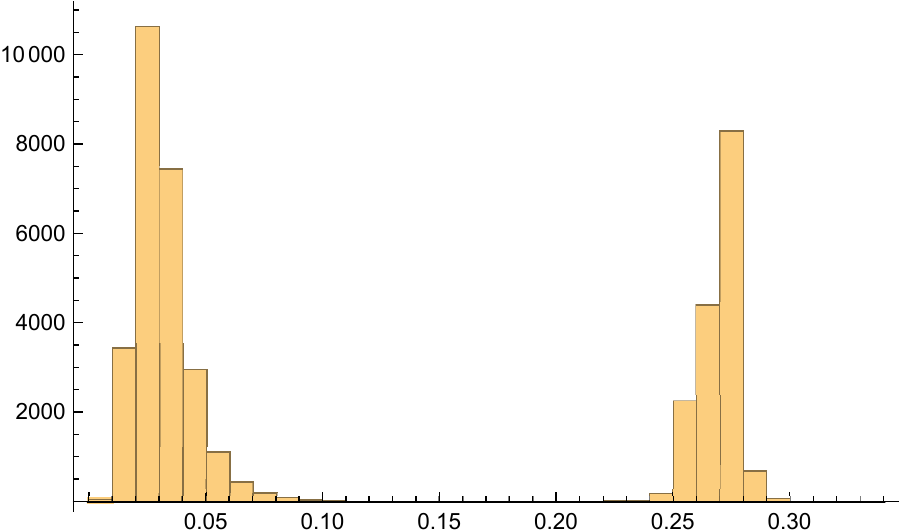}
\caption{The histograms of $\phi^j{}^2$ for $Q=Q^{SO(3)}$ with $N=16\ (L=3)$, $R=85$, 
and $\alpha=0.5$. The values 
of $\lambda$ are $10$, $10^3$, $10^5$, and $10^7$, respectively from the left to the right.}
\label{fig:transN16}
\end{figure}

In Figure~\ref{fig:transN15}, we show the results of the Monte Carlo simulation of the system \eq{eq:part}.
The value of $Q$ is taken to be $Q=Q^{SO(2)}$ with $N=15\ (M=7)$, $R=76$ (See \eq{eq:floorR}), and $\alpha=0.5$. 
Each panel shows the histogram of $\phi^j{}^2$ (no sum over $j$) for different values of $\lambda=10$, $10^3$, $10^5$, and 
$10^7$, respectively from the left to the right.
One can find the presence of a transition of the topology of the distribution: When $\lambda$ is small, the distribution forms 
one connected component, but, as $\lambda$ becomes larger, the distribution is deformed, and eventually splits into 
two connected components. 
This transition is very similar to a matrix counter part, namely, the Gross-Witten-Wadia transition \cite{Gross:1980he,Wadia:1980cp}
and the transition between
the one-cut and two-cut solutions in the matrix model \cite{Eynard:2016yaa}.
The same transition can be found for $Q=Q^{SO(3)}$ shown in Figure~\ref{fig:transN16}, and also for $Q=Q^{SO(4)}$
\cite{Kawano:2021vqc}.

\begin{figure}
\begin{center}
\includegraphics[width=7cm]{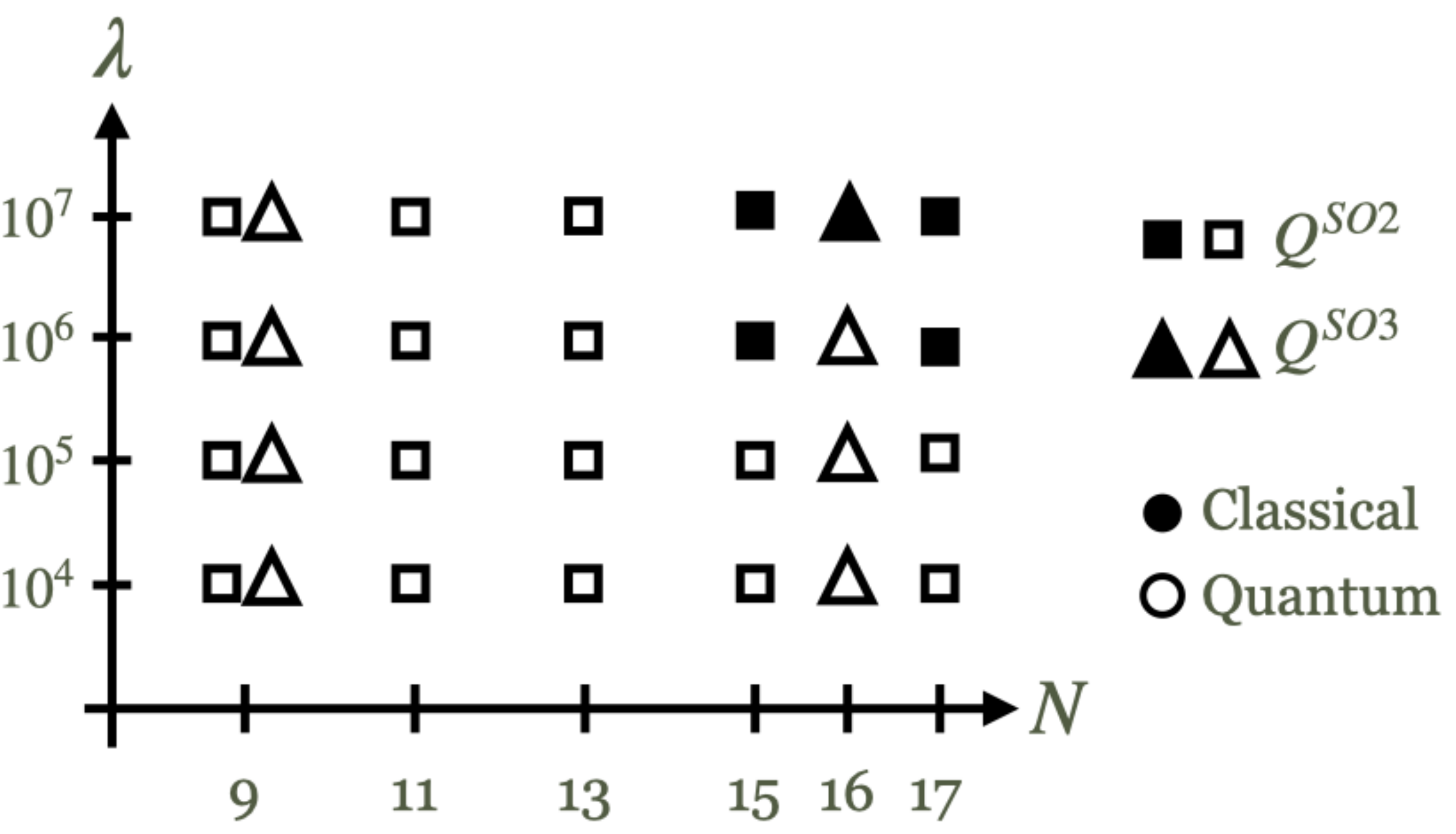}
\caption{The parameter regions of the quantum and the classical phases.}
\label{fig:region}
\end{center}
\end{figure}

Characteristics of the two phases divided by the transition can be stated as follows. 
In the phase which appears for smaller $\lambda$ can be called the quantum phase. There the fluctuation of $\phi^j{}^2$ 
is large. On the other hand, the phase which appears for larger $\lambda$ can be called the classical phase. 
In this phase, the distribution of $\phi^j{}^2$ splits into two bunches, an inner bunch and an outer one.
The fluctuation of $\phi^j{}^2$ is suppressed, because
$\phi^j{}^2$s are concentrated around one of the bunches. 
The concentration becomes stronger as $\lambda$ becomes larger.
In Figure~\ref{fig:region}, the parameter regions of the quantum and the classical phases are shown. It shows that 
the classical phase appears when $\lambda$ and $N$ are large.

\begin{figure}
\includegraphics[width=3.5cm]{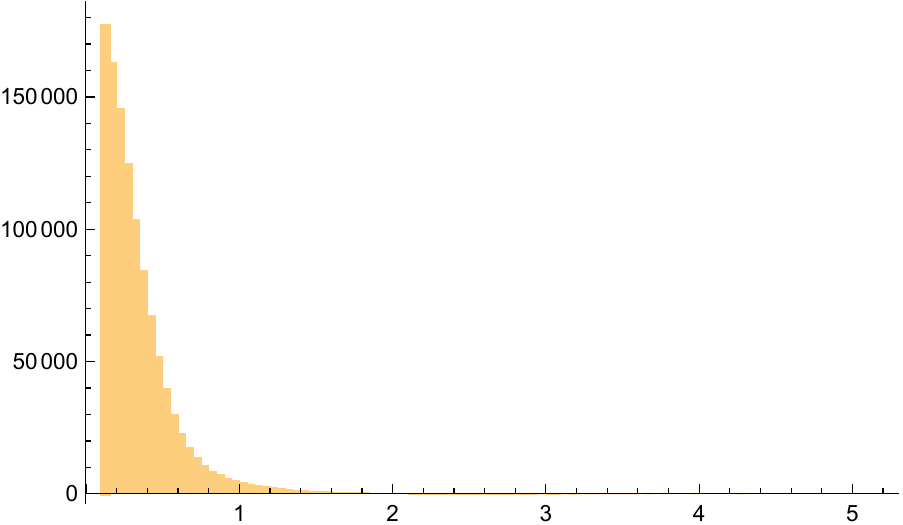}
\hfil
\includegraphics[width=3.5cm]{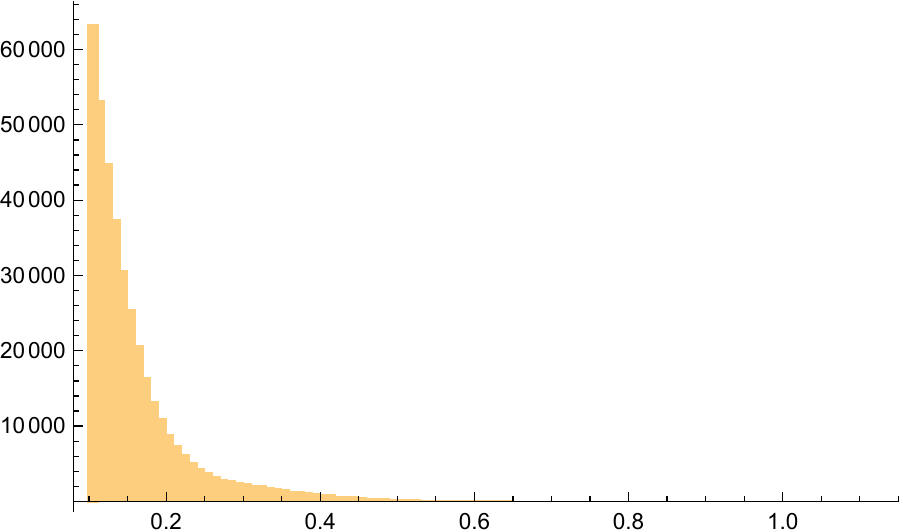}
\hfil
\includegraphics[width=3.5cm]{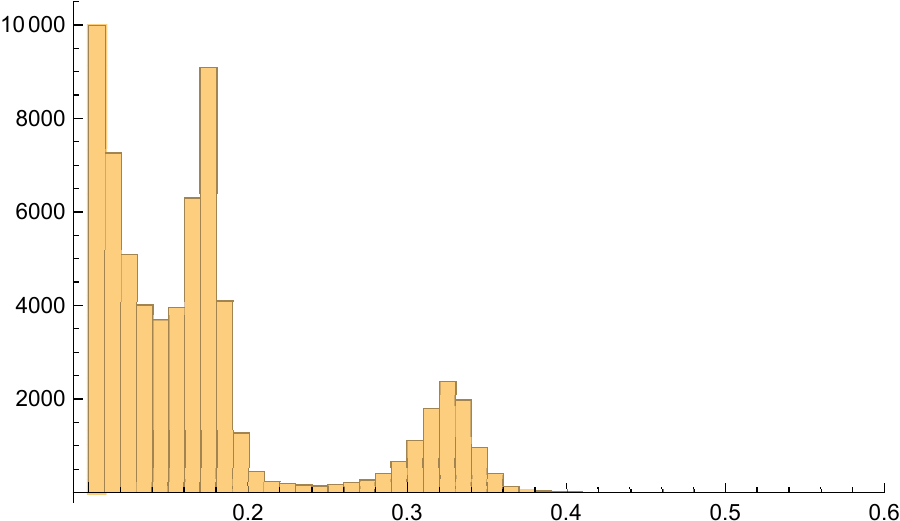}
\hfil
\includegraphics[width=3.5cm]{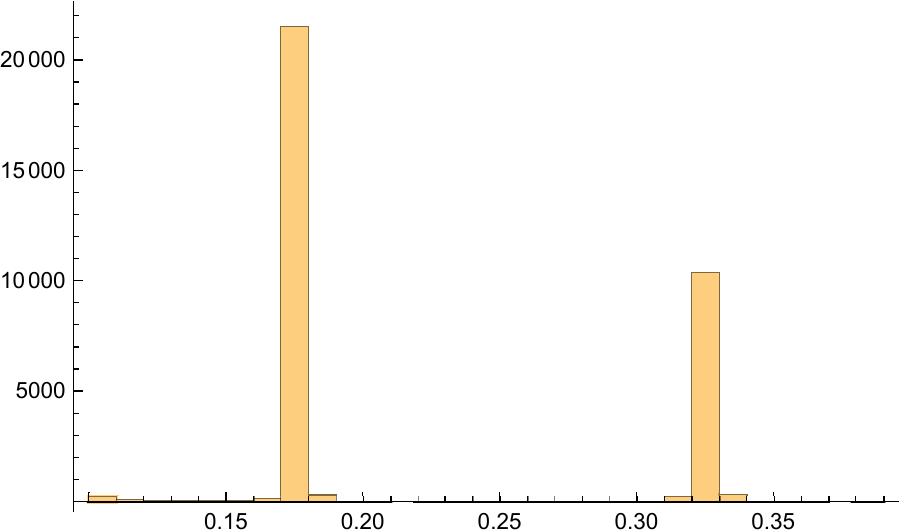}
\caption{The histograms of $\phi_a^j\phi_a^k$ for $Q=Q^{SO(2)}$ with $N=15\ (M=7)$, $R=76$, 
and $\alpha=0.5$. A cut, $\phi_a^j\phi_a^k>0.1$, is set for the histograms for convenience to ignore large 
accumulations around  $\phi_a^j\phi_a^k\sim 0$.
The values of $\lambda$ are $10$, $10^3$, $10^5$, and $10^7$, respectively from the left to the right.}
\label{fig:so2ij}
\end{figure}

In fact, the semi-classical nature of the classical phase is not restricted to
 $\phi^j{}^2$, but is also shared by the inner products, $\phi^j_a \phi_a^k$, 
as shown in Figure~\ref{fig:so2ij}.
These observations lead to the following picture of the classical phase.
When $\lambda$ becomes large, the dynamics of the system \eq{eq:part} 
requires $Q_{abc}\sim \sum_{j=1}^R \phi_a^j \phi_b^j \phi_c^j$.  In the classical phase, however,
what actually happens is that the sum is dominated by the $\phi^j$s belonging to the outer bunch, 
\beqs
 Q_{abc}\sim \sum_{\phi^j \in\, {\rm Outer}}  \phi_a^j \phi_b^j \phi_c^j,  
 \label{eq:qouter}
\eeqs
while $\phi^j\in {\rm Inner\ bunch}$ can be ignored as small contributions. 
Moreover, the inner products between $\phi^j\, \in {\rm Outer\ bunch}$ are semi-classical. 
Therefore the classical phase can be characterized by the following set of semi-classical order parameters:
\beqs
\{ \phi^i_a \phi_a^j\ | \ \phi^i,\phi^j\in {\rm Outer\ bunch} \}.
\label{eq:order}
\eeqs 
Because the pattern of \eq{eq:order} generally breaks the replica symmetry, which is the symmetry of shuffling $\phi^j$s,
the classical phase can also be rephrased as the phase with a spontaneous replica symmetry breaking.
We will see in Section~\ref{sec:emergence} that the pattern of the order parameter \eq{eq:order}
can be interpreted as emergence of discrete $S^n$ for $Q=Q^{SO(n+1)}$. 

We would like to add that the appearance of \eq{eq:qouter} in the classicl phase 
would potentially be very important in a completely different context. A tensor rank 
decomposition\footnote{A tensor rank decomposition depends generally on whether a tensor is 
decomposed into real or complex rank-one tensors and also on whether rank-one tensors are symmetric or not. 
The present one in the text is a real symmetric tenor rank decomposition.}  (often called 
CP decomposition) of a tensor is a decomposition of a tensor into rank-one tensors 
\cite{SAPM:SAPM192761164,Carroll1970,Landsberg2012,comon:hal-00923279},
\beqs
Q_{abc}=\sum_{j=1}^{\tilde R} \phi_a^j \phi^j_b \phi^j_c,
\eeqs
where $\tilde R$ is called the rank of $Q$, if it is the minimum value which realizes a decomposition.
The tensor rank decomposition is known to be quite useful in analysis of tensors generated from real-life data.
However it is known that the tensor rank decomposition is often problematic, one of the reasons of which 
comes from that there is no efficient way to determine the rank of a tensor \cite{nphard}.
Since we cannot know the rank of a tensor beforehand, it is usually necessary to repeat optimization processes, 
assuming different values of $\tilde R$ each time, to reach an appropriate (approximate) decomposition. 
Moreover it is difficult to even know whether a decomposition is appropriate or not, because 
we cannot well discriminate an over-optimized decomposition with a wrong $\tilde R$, 
which does not properly reflect the true nature of a tensor. 
On the other hand, what looks remarkable in \eq{eq:qouter} is that a reasonable value of $\tilde R$, 
namely, the number of $\phi^j \in {\rm Outer\ bunch}$, is determined automatically by the dynamics, 
even though $R$ is (quite) larger than the rank of $Q$. 
In fact it can be checked that the numbers of $\phi^j \in {\rm Outer\ bunch}$ are the same as (or very
near to) the ranks of $Q$ in our cases\footnote{This can be checked by comparing with the decomposition
by optimizations using for instance the program used in \cite{Kawano:2018pip}.}.
Though Monte Carlo simulations are generally more costly than optimizations, it would be interesting to 
pursue useful applications of this phenomenon of dynamical determination of ranks.

\section{Emergence of semi-classical spacetimes}
\label{sec:emergence}
\begin{figure}
\begin{center}
\includegraphics[width=7cm]{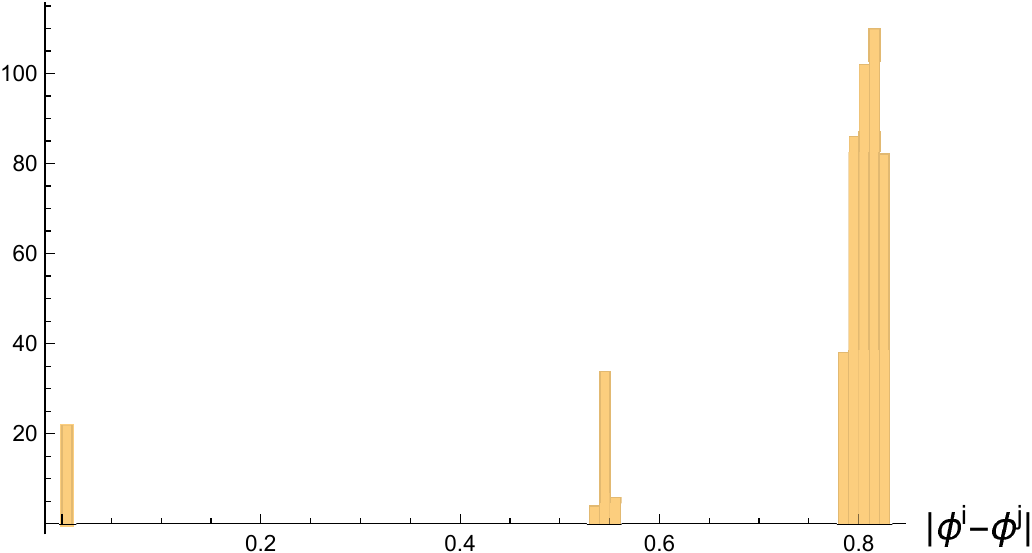}
\hfil
\includegraphics[width=3.5cm]{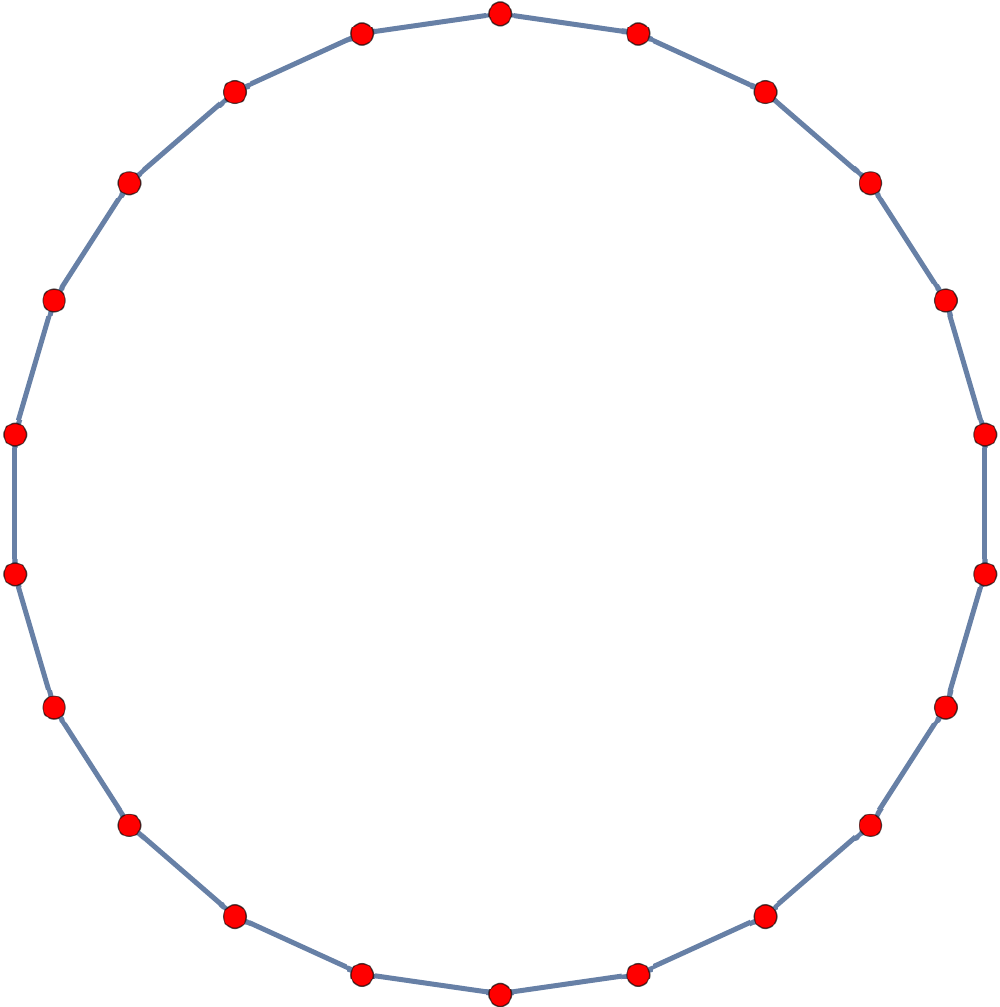}
\end{center}
\caption{Left: The histogram of distances $|\phi^j-\phi^k|$ for $Q=Q^{SO(2)}$ with $N=15\ (M=7)$, $R=76$, $\alpha=0.5$,
and $\lambda=10^7$. The nearest neighbor pairs have distances $\sim 0.55$. 
Right: The nearest neighbor pairs are connected, where each point represents each $\phi^j$ in the outer bunch.  }
\label{fig:dist15}
\end{figure}

\begin{figure}
\begin{center}
\includegraphics[width=7cm]{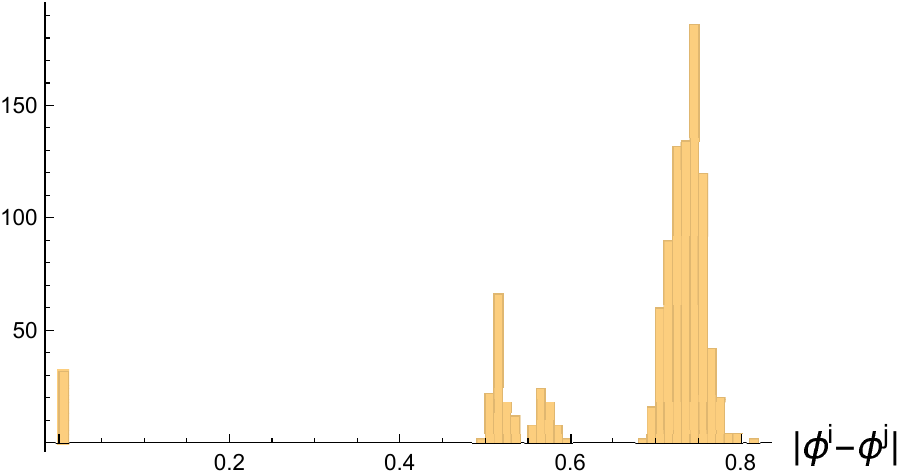}
\hfil
\includegraphics[width=3.5cm]{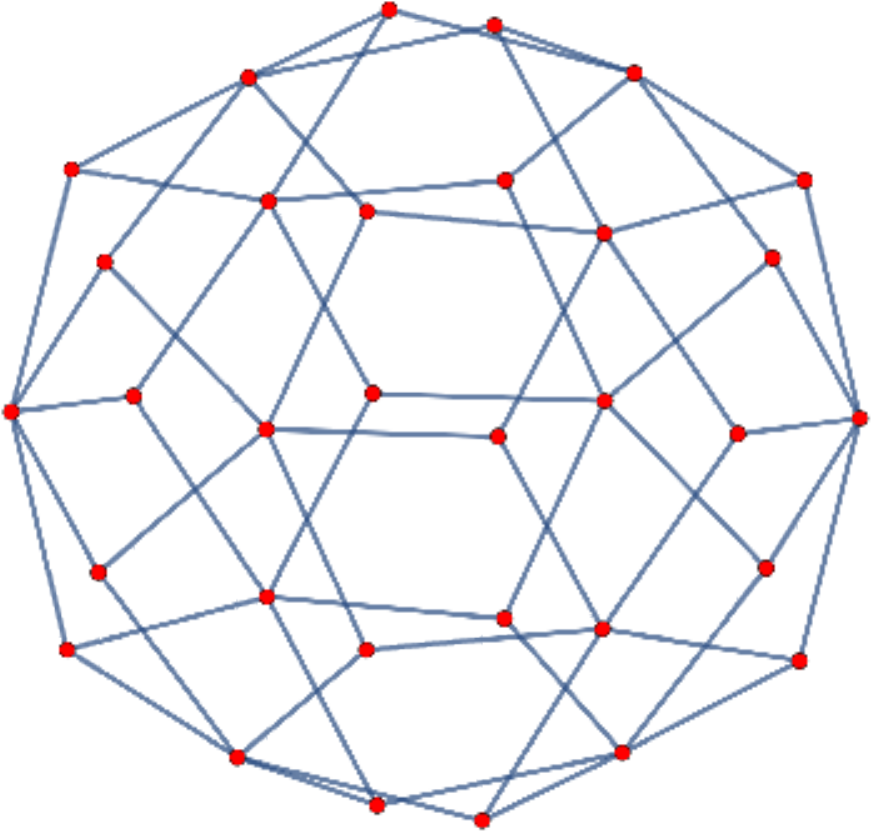}
\end{center}
\caption{Left: The histogram of distances $|\phi^j-\phi^k|$ for $Q=Q^{SO(3)}$ with $N=16\ (L=3)$, $R=85$, $\alpha=0.5$,
and $\lambda=10^7$. The nearest neighbor pairs have distances $\sim 0.52$. 
Right: The nearest neighbor pairs are connected by lines, where each point represents each $\phi^j$ in the outer bunch. }
\label{fig:dist16}
\end{figure}

In this section, we interpret the pattern of the order parameter \eq{eq:order} as emergence of a space. 
To see this, let us first note that the fluctuation is suppressed in the classical phase. Therefore, if the system is in the classical phase,
it does not lose generality to just pick up one data as a representative from a MC sequence (typically having size of $10^4\sim 10^6$). 
Then we collect the $\phi^j$s belonging to the outer bunch and compute the distances between every pair, 
$|\phi^i-\phi^j|=\sqrt{(\phi_a^i-\phi_a^j)(\phi_a^i-\phi_a^j)}$. This process determines the nearest neighbor pairs among $\phi^j$s,
and topological relations can be obtained by connecting them. 
In Figure~\ref{fig:dist15}, this is done for $Q=Q^{SO(2)}$. One can find that the topological relation clearly shows an emergence
of $S^1$. This is done for $Q=Q^{SO(3)}$ in Figure~\ref{fig:dist16}, showing an emergence of $S^2$. The $Q=Q^{SO(4)}$ 
case is not shown, because it is difficult to see $S^3$ on a two-dimensional sheet. 

The above procedure reveals the topological aspects of the emergent spaces. 
We can also study geometric aspects by defining a matrix, which has similarity with the Laplacian on an emergent
space. Let us again take one data, and take the $\phi^j$s in the outer bunch. Then let us define a matrix from
the inner products,
\beqs
A_{jk}:=\phi^j_a \phi_a^k,\ \ (j,k=1,2,\ldots,\tilde R),
\label{eq:A}
\eeqs
where $\tilde R$ is the total number of the $\phi^j$s in the outer bunch, and we have relabeled them by $\phi^j\ (j=1,2,\ldots,\tilde R)$
without loss of generality. 

This matrix \eq{eq:A} has zero or positive eigenvalues in general. In our cases treated below, $N<\tilde R$ holds and 
$A$ has a number of zero eigenvalues. Therefore, it would be more natural to consider the following matrix,
\beqs
B_{ab}:=\sum_{j=1}^{\tilde R} \phi_a^j \phi_b^j,
\eeqs 
which has the same positive eigenvalues as $A$. In our cases, the eigenvalues of $B$ are all positive, and 
it is possible to consider a matrix,
\beqs
K=-\log (B).
\eeqs

\begin{figure}
\begin{center}
\includegraphics[width=3.5cm]{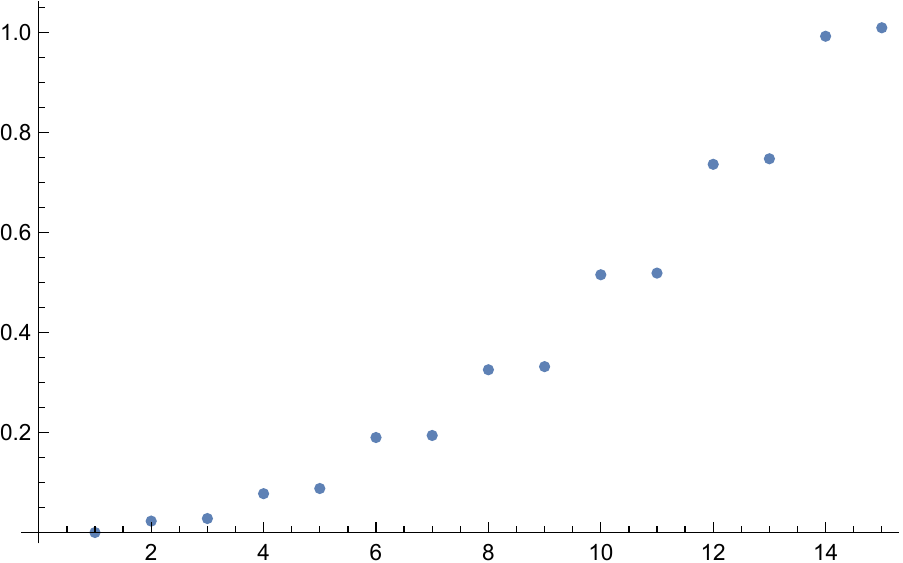}
\hfil
\includegraphics[width=3.5cm]{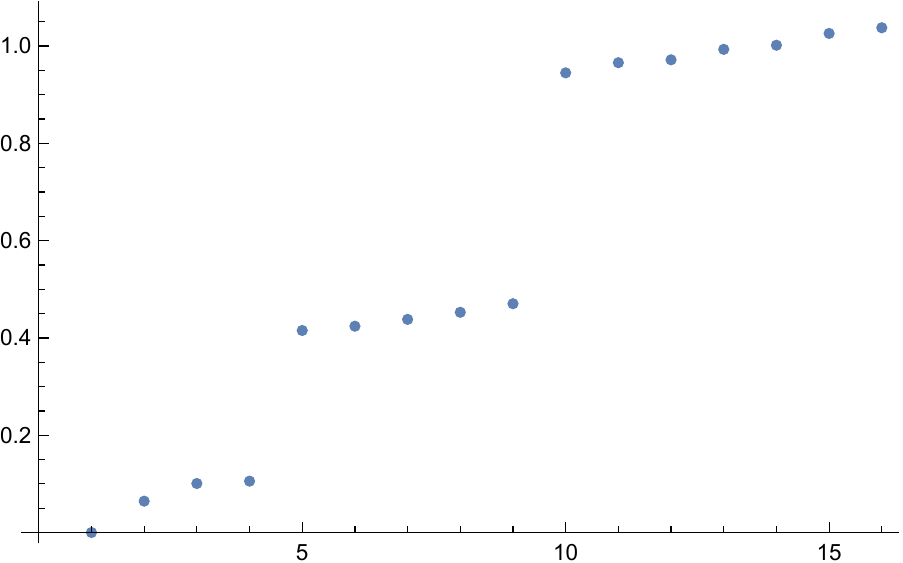}
\hfil
\includegraphics[width=3.5cm]{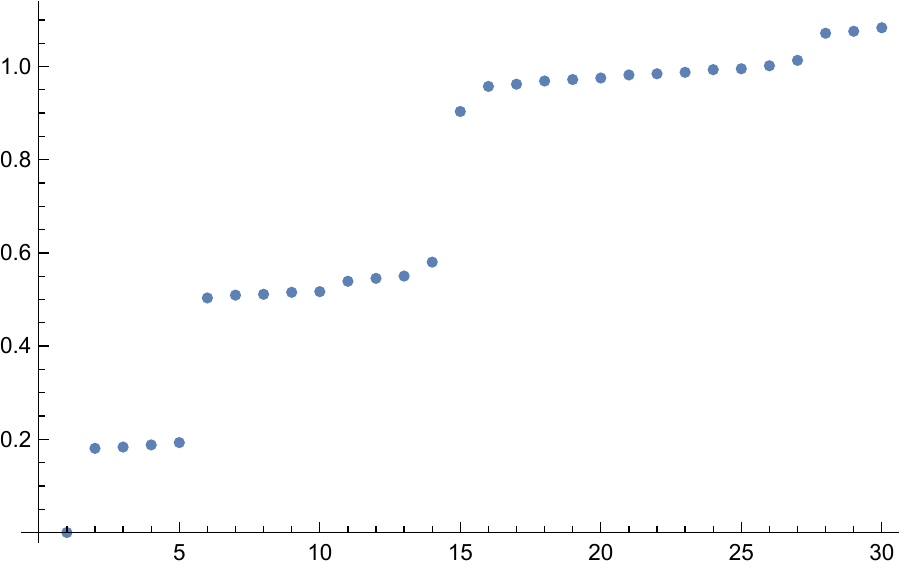}
\end{center}
\caption{Left: Spectra of $K-{\rm min}(K)$ for $Q=Q^{SO(2)}$ with $N=15\ (M=7)$, $R=76$, $\alpha=0.5$,
and $\lambda=10^7$. $\tilde R=22$ in this case. 
Middle: The same for $Q=Q^{SO(3)}$ with $N=16\ (L=3)$, $R=85$, $\alpha=0.5$,
and $\lambda=10^7$. $\tilde R=32$. 
Right: The same for $Q=Q^{SO(4)}$ with $N=30$, $R=264$, $\alpha=0.5$,
and $\lambda=10^7$. $\tilde R=78$. }
\label{fig:spec}
\end{figure}

In Figure~\ref{fig:spec}, the eigenvalues of $K-{\rm min}(K)$ are shown for $Q=Q^{SO(n+1)}\ (n=1,2,3)$. 
The eigenvalues look very much like those of Laplacians on $S^n$.
This supports that the emergent spaces are not only topologically but also geometrically $S^n$.

\section{Different values of $Q$} 
\label{sec:different}
In the previous sections, we only considered $SO(n+1)$ symmetric values of $Q$ with the particular
representations on $Q$, as in Section~\ref{sec:qc}. 
In this section, we consider perturbations of the values or change the representations to see what happens.

\begin{figure}
\begin{center}
\includegraphics[width=4.5cm]{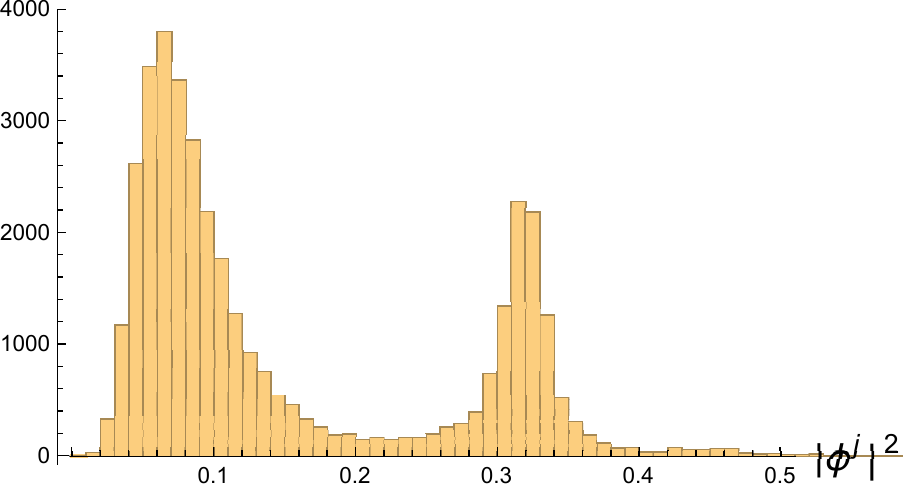}
\hfil
\includegraphics[width=4.5cm]{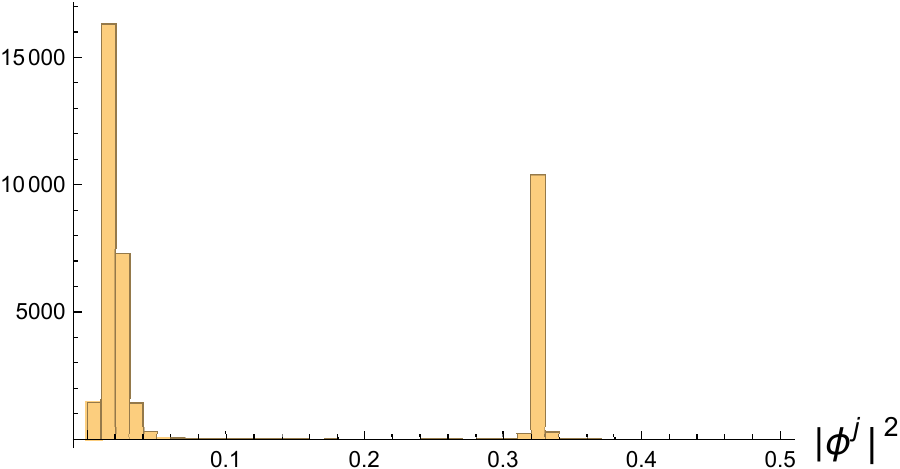}
\hfil
\includegraphics[width=4.5cm]{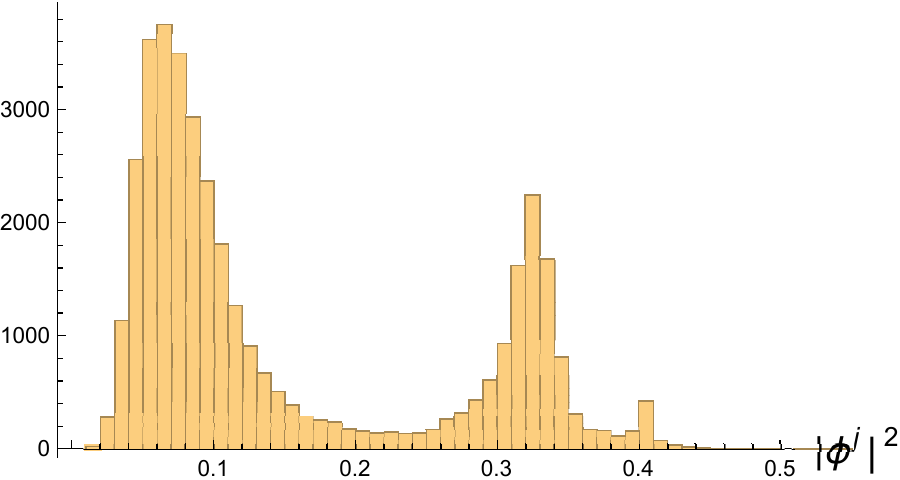}
\end{center}
\caption{
The histograms of $|\phi^j|^2$ for $z=-3/17,0,3/17$ from the left panel to the right, respectively.
The parameters are $Q^{SO(2)}$ with $N=15 \, (\Lambda=7)$, $\lambda=10^7$, and $\alpha=0.5$.
 }
\label{fig:off}
\end{figure}

The perturbations we consider have the form, 
\beqs
Q=\frac{Q^{SO(n+1)}+z\, Q^B}{\sqrt{1+z^2}},
\eeqs
where $z$ is a deformation parameter, and $Q^B\ (|Q^B|=1)$ is a tensor which breaks $SO(n+1)$ symmetry.  
Of course it is not possible to consider all the possibilities of $Q^B$, and hence we consider a perturbation given by
\beqs
Q^B_{abc}={\rm const.} \cdot \left\{ 
\begin{array}{cl}
\cos\left[0.1(a+b+c)\right]  & \hbox{if } Q^{SO(n+1)}_{abc}=0 \\
0 & \hbox{otherwise}
\end{array}
 \right.
 ,
 \label{eq:qb}
\eeqs
where const. is taken so that $|Q^B|=1$. This perturbation is so meaningless that this could represent a general aspect
under perturbations.
Figure~\ref{fig:off} shows the result for $Q^{SO(2)}$. The perturbations turn the classical phase into the quantum.

\begin{figure}
\begin{center}
\includegraphics[width=3.5cm]{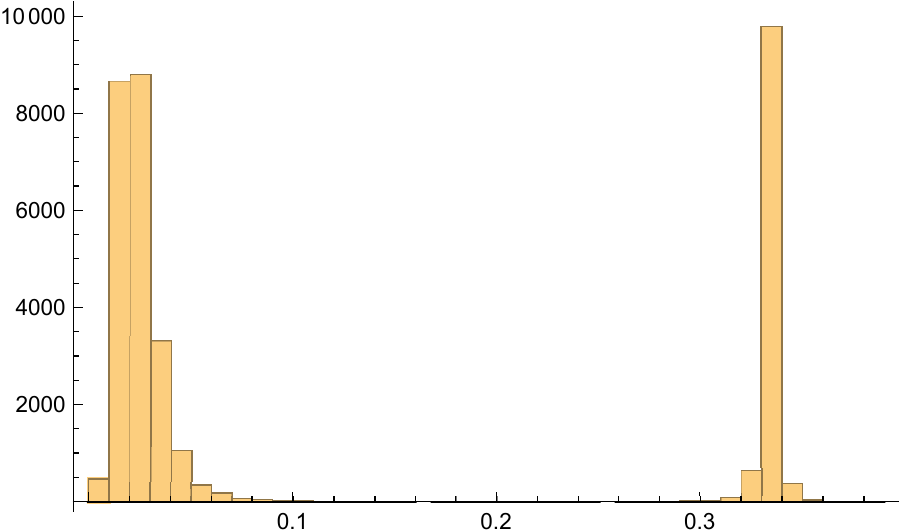}
\hfil
\includegraphics[width=3.5cm]{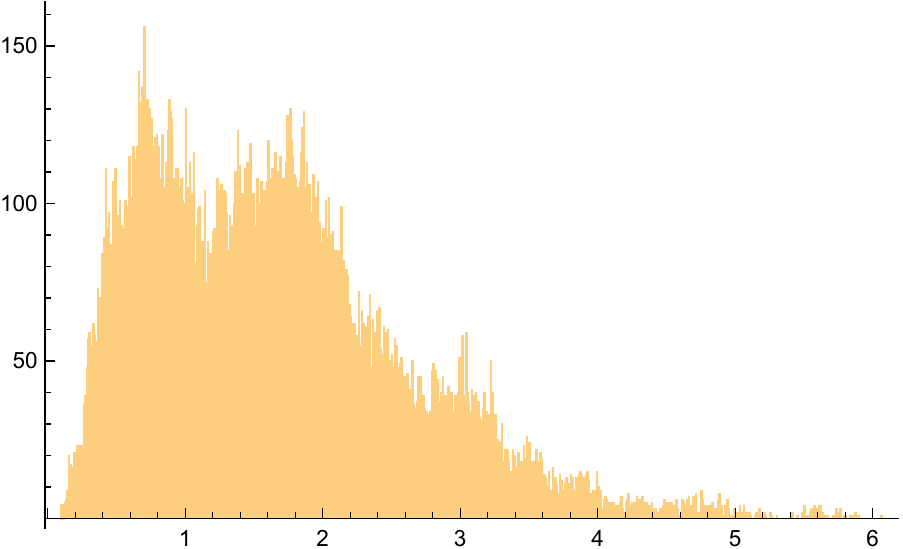}
\hfil
\includegraphics[width=3.5cm]{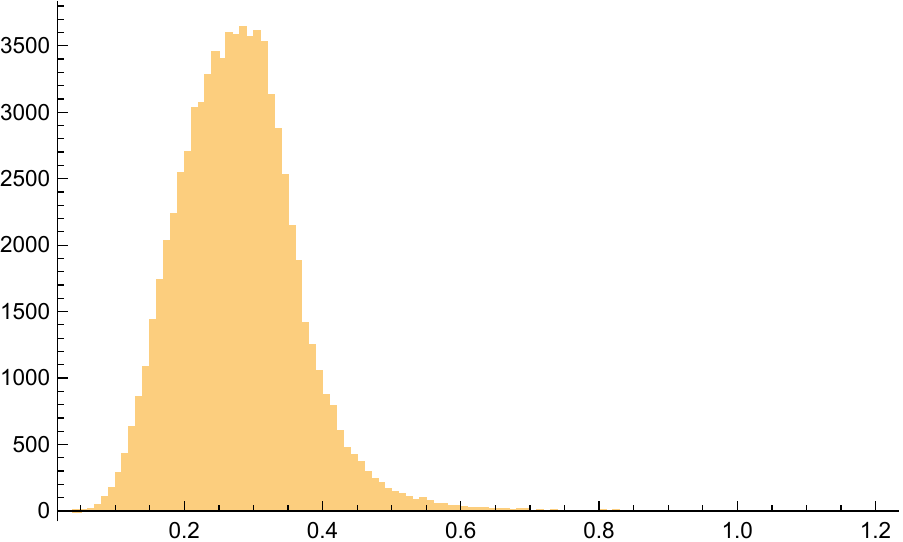}
\hfil
\includegraphics[width=3.5cm]{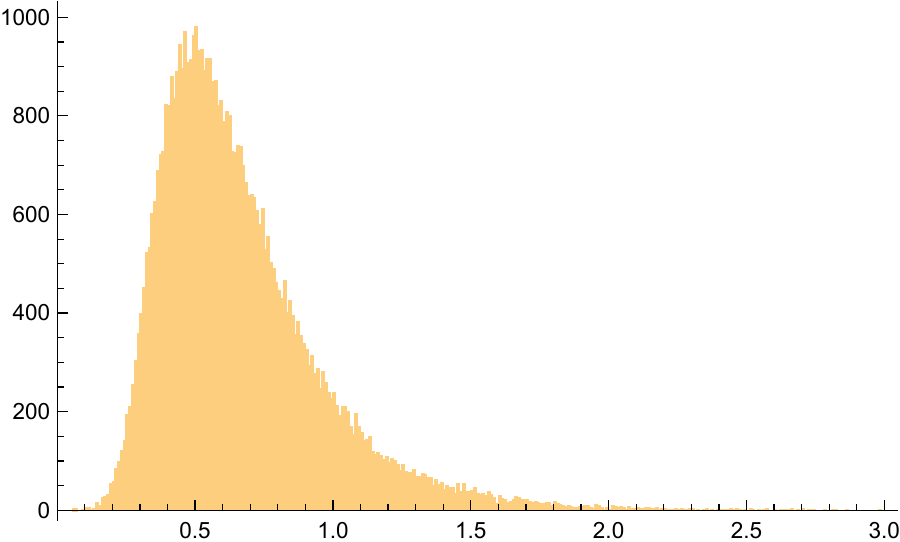}
\end{center}
\caption{
Dependence of phases on representations. Left two panels: $Q^{SO(2)}$ with $M=7$,
$\lambda=10^7$, and $\alpha=0.5$. Here $p=0$ and $p=0,1$ are dropped respectively in the left and the right.
Right two panels:  $Q^{SO(3)}$ with $L=3$, $\lambda=10^7$, and $\alpha=0.5$. 
$l=0$ and $l=0,1$ are dropped respectively in the left and the right.
 }
\label{fig:rep}
\end{figure}

One can also change the representations. In Section~\ref{sec:qc}, the representations of the $SO(n+1)$ symmetries on $Q$ 
are taken successively from the trivial representation to a representation labeled by a cutoff parameter. This is physically a 
natural choice, because the cutoff can be naturally associated to a small scale cutoff of a space. However, if we do not care 
this physical interpretation, it is free to consider any other representations. 
An option is dropping part of the irreducible representations taken in Section~\ref{sec:qc}.  
Figure~\ref{fig:rep} shows some results when this is carried out. 
Dropping some representations tend to make the classical phase less likely to appear. 

As a summary, the study of this section shows that the Lie-group symmetric values of $Q$ with the physically natural choices of representations 
taken in Section~\ref{sec:qc} make the classical phase likely to appear, which emergence of spacetimes occurs, compared to the cases 
with perturbed $Q$ or with other representations.

\section{Speculations on the evolution of spacetimes in CTM} 
\label{sec:speculations}
By performing a rescaling $\phi\rightarrow |Q|^\frac{1}{3} \phi$ in \eq{eq:psiqlam}, one can obtain the following expression which
can be used in place of \eq{eq:psiqmonte}:
\beqs
\Psi(Q,\lambda)= {\rm const.}\, |Q|^{\frac{NR}3} \, Z_{\tilde Q,\lambda |Q|^2 }\, \left \langle \prod_{j=1}^R {\rm Airy}(-|Q|^{\frac{2}{3}} \phi^j{}^2) \right \rangle_{\tilde Q,\lambda |Q|^2},
\label{eq:psiqtilde}
\eeqs
where we have introduced a normalized tensor, $\tilde Q= Q/|Q|$, satisfying $|\tilde Q|=1$.
With this expression we will speculatively discuss ``time evolution" in CTM.

The first thing we must assume is what should be taken as time in CTM. This is a non-trivial question which commonly
appears in quantizing spacetime diffeomorphism invariant (or analogous) theories \cite{Isham:1992ms}.
 From a rigid point of view, one must introduce a clock system which counts time. 
 In this paper, however, we would rather like to leave this interesting subject for future work, 
and just assume time is positively correlated with $|Q|$. 
This assumption comes from the agreement between CTM for $N=1$ 
and the mini-superspace treatment of GR, in which $Q$ is indeed proportional to the spatial volume $a^d$ of GR, where $a$ is the spatial scale
factor and $d$ is the spatial dimension. Though we do not know whether $|Q|$ is proportional to the spatial volume for the $N>1$ case as well, 
we would be able to assume that time is roughly positively correlated with $|Q|$ also for $N>1$.

To discuss time evolution under this assumption,
let us study the $|Q|$ dependence of the wave function \eq{eq:psiqtilde} with fixed $\tilde Q$. There are two effects:
One is the increase of the effective coupling $\lambda |Q|^2$, 
and the other is the increase of the coefficients in the argument of the Airy functions in \eq{eq:psiqtilde}. 

Let us first discuss the former effect. It is obvious that, when $|Q|=0$, the system is in the quantum phase.
As $|Q|$ increases, the effective coupling becomes larger, and the system may eventually encounter a transition to the classical phase.
As suggested in some examples in Section~\ref{sec:different}, it would also be possible that there are no transitions however 
large $|Q|$ becomes.
The results 
there also suggest that the transition occurs at relatively smaller values of $|Q|$ for Lie-group symmetric $\tilde Q$.

Let us next discuss the latter effect. 
In fact the latter effect is more interesting, reflecting a quantum aspect of CTM. 
Let us consider the quantum phase, 
where the fluctuations of $\phi^j{}^2$s are large, as shown in Section~\ref{sec:qc}.
When $|Q|$ becomes larger, the fluctuations are enhanced in the argument of 
the Airy functions in \eq{eq:psiqtilde}. This means that, when $|Q|$ becomes larger, the wave function is more suppressed because 
the cancellations dominate more (Recall that the Airy function is oscillatory as in Figure~\ref{fig:airy}). 
Therefore, the wave function will be more suppressed as $|Q|$ becomes larger in the quantum phase than 
in the classical phase.  Some explicit cases of the suppression can be found in the previous paper \cite{Kawano:2021vqc}, 
where one can find a rather strong suppression of the expectation value, $\langle \prod_{j} {\rm Airy}\rangle$, in \eq{eq:psiqtilde}. 

\begin{figure}
\begin{center}
\includegraphics[width=3cm]{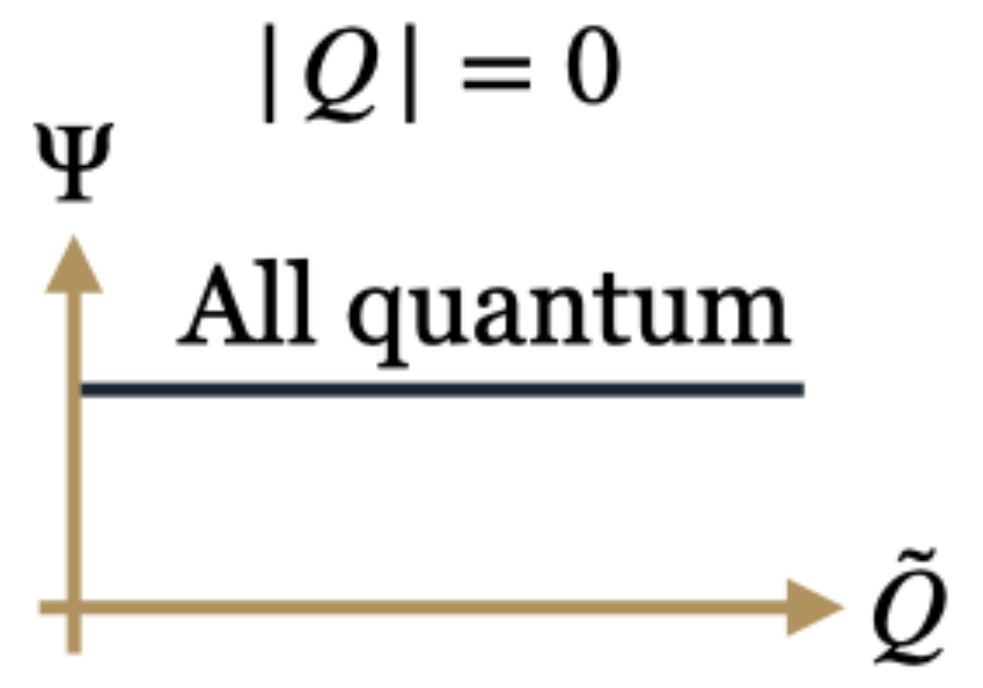}
\hfil
\includegraphics[width=3.5cm]{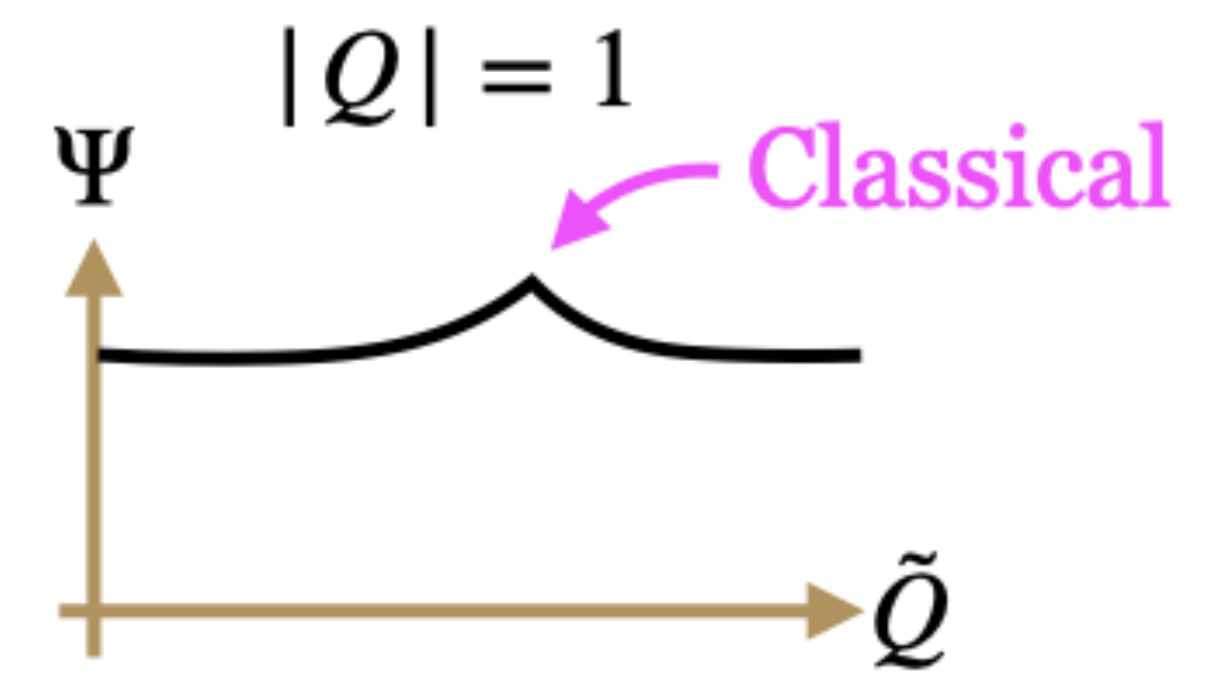}
\hfil
\includegraphics[width=3.5cm]{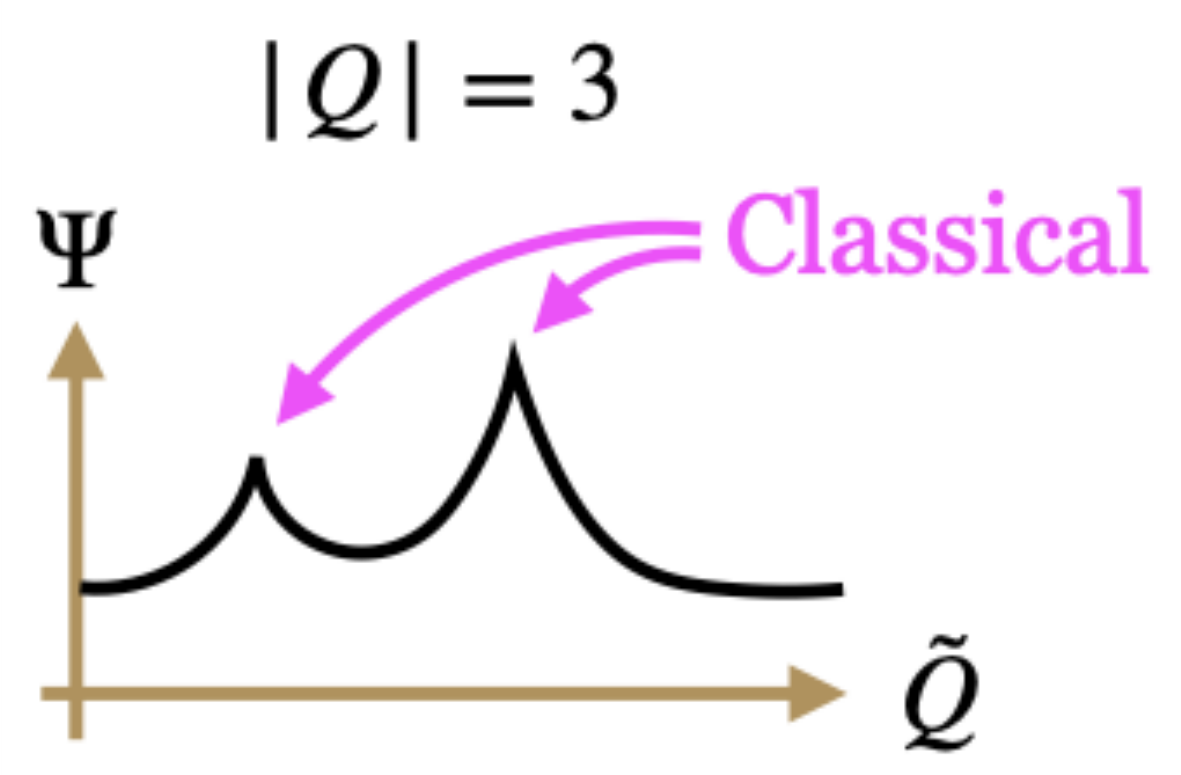}
\hfil
\includegraphics[width=3.5cm]{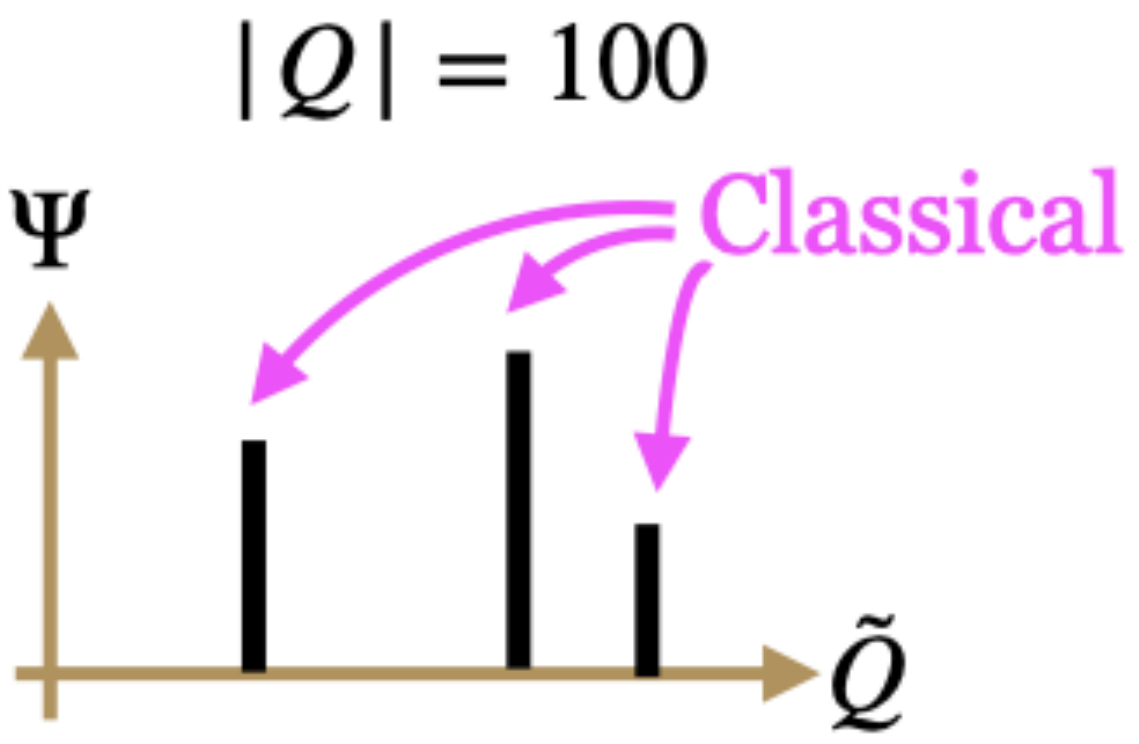}
\caption{The illustration of the  ``time evolution" in CTM. Only $\tilde Q$s in the classical phase remain, as $|Q|$ becomes larger. 
The values of $|Q|$ in the figures are taken arbitrary for illustration.}
\label{fig:time}
\end{center}
\end{figure}

The above consideration leads to the following speculative picture about the ``time evolution" in CTM, as illustrated 
in Figure~\ref{fig:time}. When $|Q|=0$, the system is in 
the quantum phase whatever values $\tilde Q$ takes. Then, as $|Q|$ develops, the systems at some $\tilde Q$s encounter 
transitions to the classical phase. Since the wave function in the quantum phase is more suppressed as $|Q|$ increases,
the wave function gradually develop a structure in which only $\tilde Q$s in the classical phase remain (See the rightmost panel).
The amplitudes of the wave function at such $\tilde Q$s will be larger, 
as the system encounters the transition at smaller values of $|Q|$ (Compare the right three panels).

\section{Summary and future prospects}
In this paper, we have reviewed some preliminary results of the study of the wave function of CTM \cite{Sasakura:2011sq}
in $Q$-representation, which was 
performed by the Hamiltonian Monte Carlo simulation with the reweighting method \cite{Sasakura:2021lub,
Kawano:2021vqc}. 
The most important result was the discovery of the classical phase, in which classical spacetimes emerge.
More concretely, we have demonstrated the emergence of discrete $n$-dimensional spheres for $SO(n+1)$-invariant values of $Q$ with
physically natural choices of representations. Based on the results, we have speculated how spacetimes evolve in CTM: 
Initially the system is totally in the quantum phase with no classical structures, but, as time develops (namely, $|Q|$ becomes larger), 
some configurations (described by $\tilde Q$) encounter the transition to the classical phase to generate classical spaces, 
and only these configurations eventually remain, since all the other 
configurations in the quantum phase are eventually suppressed to vanish.  

A next physically important question arising from this paper will be whether the classical spacetimes which emerge in CTM follow 
the equation of motion of GR. Since CTM enjoys similar structure as the Hamiltonian formalism of GR (namely, the ADM formalism),
this is highly expected. In fact, a few connections between CTM and GR have already been known 
\cite{Sasakura:2014gia,Sasakura:2015pxa,Chen:2016ate}: In particular in 
\cite{Chen:2016ate}, it has been 
shown that the classical equation of motion of CTM in a formal continuum limit agrees with a Hamilton-Jacobi equation of GR 
with a Hamilton's principal function of a local form. It would be interesting to perform similar analysis for the configurations at 
wave function peaks with emergence of classical spacetimes.

Another interesting question would be to more thoroughly study the transition between the quantum and the classical phases. 
The transition has the form of a topological change of distributions, which has some similarity with the matrix counter parts,
the transition between the one-cut and two-cut solutions in the matrix model \cite{Eynard:2016yaa} 
or the Gross-Witten-Wadia transition \cite{Gross:1980he,Wadia:1980cp}.
However, we do not currently know how sensible the similarity is, since properties of the transition 
are largely unknown, such as the thermodynamic limit to make it a sharp phase transition, 
the order of the transition, and so on.
Moreover the transition was found for the system defined by the real part of the wave function 
(in the reweighting method), but,
from the point of view of CTM, the transition should be studied for the full system including the complex part. 
It would also be exciting to explore the analogy between spacetimes and glasses, 
since the expression of the wave function has good similarities with the spherical $p$-spin model for spin 
glasses \cite{pspin,pedestrians}. In fact, the transition to the glassy phase is described by the replica symmetry breaking, 
which is nothing but what characterizes the classical phase, in which spacetimes emerge in CTM.
 
Another more ambitious direction of study is to reveal phenomenological aspects of CTM. It would be highly interesting
if one can mange to make precise the time evolution of the emergent spacetimes, and compute some 
observables which can be compared with astrophysical observations, such as primordial fluctuations.
This would approve/disapprove CTM as a model of the Universe.  

Last but not least, we would like to point out a potential importance of the classical phase in terms of tensor rank decomposition.
Tensor rank decomposition \cite{SAPM:SAPM192761164,Carroll1970,Landsberg2012,comon:hal-00923279}
is known to be an effective method to extract information from tensors generated from real-life data,
but there are some technical issues. One is that, since there is no efficient way to know the 
rank of a tensor beforehand \cite{nphard}, 
one usually has to repeat the optimization process of tensor rank decomposition, 
changing trial values of ranks each time, until one gets a satisfactory decomposition.
This is not only time-consuming but also means that there are ambiguities of extracted information,
since the condition of a satisfactory tensor rank decomposition is ambiguous.
On the other hand,
what is interesting in the classical phase is that a tensor rank decomposition is automatically performed with a dynamically determined rank.
Moreover the number is exactly or very near to the rank of a tensor, at least in our cases. 
Though Monte Carlo simulations are generally much costly compared to optimizations, 
it would be interesting to pursue applications of this dynamical method of tensor rank decomposition 
to tensors from real-life. 

\acknowledgments
The work of N.S. is supported in part by JSPS KAKENHI Grant No.19K03825.


\begin{thebibliography}{99}

\bibitem{Garay:1994en}
L.~J.~Garay,
``Quantum gravity and minimum length,''
Int. J. Mod. Phys. A \textbf{10}, 145-166 (1995)
doi:10.1142/S0217751X95000085
[arXiv:gr-qc/9403008 [gr-qc]].

\bibitem{Ambjorn:1990ge}
J.~Ambjorn, B.~Durhuus, and T.~Jonsson, ``{Three-dimensional simplicial quantum
  gravity and generalized matrix models},''
\href{http://dx.doi.org/10.1142/S0217732391001184}{{\em Mod. Phys. Lett.}
  {\bfseries A06} (1991) 1133--1146}.

\bibitem{Sasakura:1990fs}
N.~Sasakura, ``{Tensor model for gravity and orientability of manifold},''
\href{http://dx.doi.org/10.1142/S0217732391003055}{{\em Mod. Phys. Lett.}
  {\bfseries A06} (1991) 2613--2624}.

\bibitem{Godfrey:1990dt}
N.~Godfrey and M.~Gross, ``{Simplicial quantum gravity in more than
  two-dimensions},''
\href{http://dx.doi.org/10.1103/PhysRevD.43.1749}{{\em Phys. Rev.} {\bfseries
  D43} (1991) R1749--1753}.

\bibitem{Gurau:2009tw} 
  R.~Gurau,
  ``Colored Group Field Theory,''
  Commun.\ Math.\ Phys.\  {\bf 304}, 69 (2011)
  doi:10.1007/s00220-011-1226-9
  [arXiv:0907.2582 [hep-th]].
  
\bibitem{Bonzom:2011zz}
V.~Bonzom, R.~Gurau, A.~Riello and V.~Rivasseau,
``Critical behavior of colored tensor models in the large N limit,''
Nucl. Phys. B \textbf{853}, 174-195 (2011)
doi:10.1016/j.nuclphysb.2011.07.022
[arXiv:1105.3122 [hep-th]].
  
\bibitem{Gurau:2011xp}
R.~Gurau and J.~P.~Ryan,
``Colored Tensor Models - a review,''
SIGMA \textbf{8}, 020 (2012)
doi:10.3842/SIGMA.2012.020
[arXiv:1109.4812 [hep-th]].

\bibitem{Ambjorn:2004qm}
J.~Ambjorn, J.~Jurkiewicz and R.~Loll,
``Emergence of a 4-D world from causal quantum gravity,''
Phys. Rev. Lett. \textbf{93}, 131301 (2004)
doi:10.1103/PhysRevLett.93.131301
[arXiv:hep-th/0404156 [hep-th]].

\bibitem{Sasakura:2011sq}
N.~Sasakura,
``Canonical tensor models with local time,''
Int. J. Mod. Phys. A \textbf{27}, 1250020 (2012)
doi:10.1142/S0217751X12500200
[arXiv:1111.2790 [hep-th]].

\bibitem{Sasakura:2012fb}
N.~Sasakura,
``Uniqueness of canonical tensor model with local time,''
Int. J. Mod. Phys. A \textbf{27}, 1250096 (2012)
doi:10.1142/S0217751X12500960
[arXiv:1203.0421 [hep-th]].

\bibitem{Arnowitt:1962hi}
R.~L.~Arnowitt, S.~Deser and C.~W.~Misner,
``The Dynamics of general relativity,''
Gen. Rel. Grav. \textbf{40}, 1997-2027 (2008)
doi:10.1007/s10714-008-0661-1
[arXiv:gr-qc/0405109 [gr-qc]].

\bibitem{Sasakura:2014gia}
N.~Sasakura and Y.~Sato,
``Interpreting canonical tensor model in minisuperspace,''
Phys. Lett. B \textbf{732}, 32-35 (2014)
doi:10.1016/j.physletb.2014.03.006
[arXiv:1401.2062 [hep-th]].

\bibitem{Sasakura:2015pxa}
N.~Sasakura and Y.~Sato,
``Constraint algebra of general relativity from a formal continuum limit of canonical tensor model,''
JHEP \textbf{10}, 109 (2015)
doi:10.1007/JHEP10(2015)109
[arXiv:1506.04872 [hep-th]].

\bibitem{Chen:2016ate}
H.~Chen, N.~Sasakura and Y.~Sato,
``Equation of motion of canonical tensor model and Hamilton-Jacobi equation of general relativity,''
Phys. Rev. D \textbf{95}, no.6, 066008 (2017)
doi:10.1103/PhysRevD.95.066008
[arXiv:1609.01946 [hep-th]].

\bibitem{Sasakura:2013wza}
N.~Sasakura,
``Quantum canonical tensor model and an exact wave function,''
Int. J. Mod. Phys. A \textbf{28}, 1350111 (2013)
doi:10.1142/S0217751X1350111X
[arXiv:1305.6389 [hep-th]].

\bibitem{Narain:2014cya}
G.~Narain, N.~Sasakura and Y.~Sato,
``Physical states in the canonical tensor model from the perspective of random tensor networks,''
JHEP \textbf{01}, 010 (2015)
doi:10.1007/JHEP01(2015)010
[arXiv:1410.2683 [hep-th]].

\bibitem{Obster:2017dhx}
D.~Obster and N.~Sasakura,
``Emergent symmetries in the canonical tensor model,''
PTEP \textbf{2018}, no.4, 043A01 (2018)
doi:10.1093/ptep/pty038
[arXiv:1710.07449 [hep-th]].

\bibitem{Obster:2017pdq}
D.~Obster and N.~Sasakura,
``Symmetric configurations highlighted by collective quantum coherence,''
Eur. Phys. J. C \textbf{77}, no.11, 783 (2017)
doi:10.1140/epjc/s10052-017-5355-y
[arXiv:1704.02113 [hep-th]].

\bibitem{Sasakura:2021lub}
N.~Sasakura,
``Phase profile of the wave function of canonical tensor model and emergence of large spacetimes,''
Int. J. Mod. Phys. A \textbf{36}, no.29, 2150222 (2021)
doi:10.1142/S0217751X21502225
[arXiv:2104.11845 [hep-th]].

\bibitem{Kawano:2021vqc}
T.~Kawano and N.~Sasakura,
``Emergence of Lie group symmetric classical spacetimes in canonical tensor model,''
[arXiv:2109.09896 [hep-th]], to be published in PTEP.

\bibitem{Obster:2020vfo}
D.~Obster and N.~Sasakura,
``Phases of a matrix model with non-pairwise index contractions,''
PTEP \textbf{2020}, no.7, 073B06 (2020)
doi:10.1093/ptep/ptaa085
[arXiv:2004.03152 [hep-th]].

\bibitem{Sasakura:2019hql}
N.~Sasakura and S.~Takeuchi,
``Numerical and analytical analyses of a matrix model with non-pairwise contracted indices,''
Eur. Phys. J. C \textbf{80}, no.2, 118 (2020)
doi:10.1140/epjc/s10052-019-7591-9
[arXiv:1907.06137 [hep-th]].

\bibitem{Witten:2010cx}
E.~Witten,
``Analytic Continuation Of Chern-Simons Theory,''
AMS/IP Stud. Adv. Math. \textbf{50}, 347-446 (2011)
[arXiv:1001.2933 [hep-th]].

\bibitem{Berger:2019odf}
C.~E.~Berger, L.~Rammelm\"uller, A.~C.~Loheac, F.~Ehmann, J.~Braun and J.~E.~Drut,
``Complex Langevin and other approaches to the sign problem in quantum many-body physics,''
Phys. Rept. \textbf{892}, 1-54 (2021)
doi:10.1016/j.physrep.2020.09.002
[arXiv:1907.10183 [cond-mat.quant-gas]].

\bibitem{Obster:2021xtb}
D.~Obster and N.~Sasakura,
``Counting Tensor Rank Decompositions,''
Universe \textbf{7}, no.8, 302 (2021)
doi:10.3390/universe7080302
[arXiv:2107.10237 [gr-qc]].

\bibitem{Neal(2011)} 
R.~Neal, Handbook of Markov Chain Monte Carlo, 113 (2011). 
doi:10.1201/b10905
[arXiv:1206.1901 [stat.CO]].

\bibitem{boost} \url{https://www.boost.org}

\bibitem{Gross:1980he}
D.~J.~Gross and E.~Witten,
``Possible Third Order Phase Transition in the Large N Lattice Gauge Theory,''
Phys. Rev. D \textbf{21}, 446-453 (1980)
doi:10.1103/PhysRevD.21.446

\bibitem{Wadia:1980cp}
S.~R.~Wadia,
``$N$ = Infinity Phase Transition in a Class of Exactly Soluble Model Lattice Gauge Theories,''
Phys. Lett. B \textbf{93}, 403-410 (1980)
doi:10.1016/0370-2693(80)90353-6

\bibitem{Eynard:2016yaa}
B.~Eynard,
``Counting Surfaces,''
Prog.~Math.~Phys. 70 (2016)
doi:10.1007/978-3-7643-8797-6

\bibitem{SAPM:SAPM192761164}
F.~L. Hitchcock, ``The expression of a tensor or a polyadic as a sum of
  products,'' \href{http://dx.doi.org/10.1002/sapm192761164}{{\em Journal of
  Mathematics and Physics} {\bfseries 6} no.~1-4, (1927) 164--189}.
  \url{http://dx.doi.org/10.1002/sapm192761164}.

\bibitem{Carroll1970}
J.~D. Carroll and J.-J. Chang, ``Analysis of individual differences in
  multidimensional scaling via an n-way generalization of ``eckart-young''
  decomposition,'' \href{http://dx.doi.org/10.1007/BF02310791}{{\em
  Psychometrika} {\bfseries 35} no.~3, (Sep, 1970) 283--319}.
  \url{https://doi.org/10.1007/BF02310791}.

\bibitem{Landsberg2012}
{Landsberg, J. M.}, {\em {Tensors: Geometry and Applications}}.
\newblock {American Mathematical Society, Providence}, {2012}.

\bibitem{comon:hal-00923279}
P.~Comon, ``{Tensors: a Brief Introduction},''
  \href{http://dx.doi.org/10.1109/MSP.2014.2298533}{{\em {IEEE Signal
  Processing Magazine}} {\bfseries 31} no.~3, (May, 2014) 44--53}.
  \url{https://hal.archives-ouvertes.fr/hal-00923279}.
  
 \bibitem{nphard}  
C.~J.~Hillar and L.~Lim, ``Most tensor problems are NP-Hard", Journal of the ACM. 60 (6) (2013): 1-39. arXiv:0911.1393. doi:10.1145/2512329

\bibitem{Kawano:2018pip}
T.~Kawano, D.~Obster and N.~Sasakura,
``Canonical tensor model through data analysis: Dimensions, topologies, and geometries,''
Phys. Rev. D \textbf{97}, no.12, 124061 (2018)
doi:10.1103/PhysRevD.97.124061
[arXiv:1805.04800 [hep-th]].

\bibitem{Isham:1992ms}
C.~J.~Isham,
``Canonical quantum gravity and the problem of time,''
NATO Sci. Ser. C \textbf{409}, 157-287 (1993)
[arXiv:gr-qc/9210011 [gr-qc]].

\bibitem{pspin}
A.~Crisanti and H.-J.~Sommers, ``The spherical p-spin interaction spin glass model: the statics," 
Z.~Phys. B \textbf{87}, 341 (1992)

\bibitem{pedestrians}
T.~Castellani and A.~Cavagna, ``Spin-glass theory for pedestrians'', 
J.~Stat.~Mech.: Theo.~Exp. {\bf 2005}, 05012
[arXiv: cond-mat/0505032].  


\end{thebibliography}
\end{document}